\documentclass[twoside,twocolumn,9pt]{article}
\usepackage{extsizes}
\usepackage{amssymb}
\usepackage[super,sort&compress,comma]{natbib} 
\usepackage[version=3]{mhchem}
\usepackage[left=1.5cm, right=1.5cm, top=1.785cm, bottom=2.0cm]{geometry}
\usepackage{balance}
\usepackage{calrsfs}
\DeclareMathAlphabet{\pazocal}{OMS}{zplm}{m}{n}
\usepackage{mathptmx}
\usepackage{sectsty}
\usepackage{graphicx} 
\usepackage{lastpage}
\usepackage[format=plain,justification=justified,singlelinecheck=false,font={stretch=1.125,small,sf},labelfont=bf,labelsep=space]{caption}
\usepackage{float}
\usepackage{fancyhdr}
\usepackage{fnpos}
\usepackage[english]{babel}
\addto{\captionsenglish}{%
  
}
\usepackage{array}
\usepackage{droidsans}
\usepackage{charter}
\usepackage[T1]{fontenc}
\usepackage[usenames,dvipsnames]{xcolor}
\usepackage{setspace}
\usepackage[compact]{titlesec}
\usepackage{hyperref}

\usepackage{epstopdf}
\usepackage{graphicx}
\usepackage{dcolumn}
\usepackage{bm}
\usepackage{comment}
\usepackage{xcolor}
\usepackage{movie15}
\usepackage{caption}
\usepackage{subcaption}
\usepackage{siunitx}
\usepackage{soul}
\newcommand{\bi}[1]{\boldsymbol{#1}}

\newcommand{\at}[1]{\bigg|_{#1}}

\newcommand{\X}{\mathbf{X}}
\newcommand{\Y}{\mathbf{Y}}
\newcommand{\Z}{\mathbf{Z}}

\newcommand{\vi}{\bi{v}}

\newcommand{\x}{\bi{x}}

\newcommand{\y}{\bi{y}}

\newcommand{\n}{\hat{\bf n}}

\newcommand{\bphi}{\bar \phi}
\newcommand{\Dbp}{\Delta\bar \phi}

\newcommand{\bgamma}{\bi{\gamma}}

\newcommand{\J}{\mathbf{J}}

\newcommand{\comtg}{\bar{\mu}}

\newcommand{\tilU}{\Pi}

\newcommand{\F}{\pazocal{F}}
\newcommand{\Fdens}{F}
\newcommand{\tilF}{\F'}
\newcommand{\tilFdens}{\Fdens'}
\newcommand{\bK}{\bar{K}}
\newcommand{\bD}{\bar{D}}
\newcommand{\bDT}{\bar{D}_T}
\newcommand{\Heav}{\mathcal{H}}
\newcommand{\Texp}{T_1}

\newcommand{\ST}{\Sigma_{T}}
\newcommand{\Din}{D_{\rm +}}
\newcommand{\Dout}{D_{\rm -}}
\definecolor{cream}{RGB}{222,217,201}

\begin{document}

\pagestyle{fancy}
\thispagestyle{plain}
\fancypagestyle{plain}{
\renewcommand{\headrulewidth}{0pt}
}

\makeFNbottom
\makeatletter
\renewcommand\LARGE{\@setfontsize\LARGE{15pt}{17}}
\renewcommand\Large{\@setfontsize\Large{12pt}{14}}
\renewcommand\large{\@setfontsize\large{10pt}{12}}
\renewcommand\footnotesize{\@setfontsize\footnotesize{7pt}{10}}
\makeatother

\renewcommand{\thefootnote}{\fnsymbol{footnote}}
\renewcommand\footnoterule{\vspace*{1pt}%
\color{cream}\hrule width 3.5in height 0.4pt \color{black}\vspace*{5pt}} 
\setcounter{secnumdepth}{5}

\makeatletter 
\renewcommand\@biblabel[1]{#1}            
\renewcommand\@makefntext[1]%
{\noindent\makebox[0pt][r]{\@thefnmark\,}#1}
\makeatother 
\renewcommand{\figurename}{\small{Fig.}~}
\sectionfont{\sffamily\Large}
\subsectionfont{\normalsize}
\setstretch{1.125} 
\setlength{\skip\footins}{0.8cm}
\setlength{\footnotesep}{0.25cm}
\setlength{\jot}{10pt}
\titlespacing*{\section}{0pt}{4pt}{4pt}
\titlespacing*{\subsection}{0pt}{15pt}{1pt}

\fancyfoot{}
\fancyfoot[LO,RE]{\vspace{-7.1pt}\includegraphics[height=9pt]{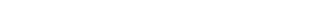}}
\fancyfoot[CO]{\vspace{-7.1pt}\hspace{13.2cm}\includegraphics{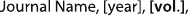}}
\fancyfoot[CE]{\vspace{-7.2pt}\hspace{-14.2cm}\includegraphics{head_foot/RF}}
\fancyfoot[RO]{\footnotesize{\sffamily{1--\pageref{LastPage} ~\textbar  \hspace{2pt}\thepage}}}
\fancyfoot[LE]{\footnotesize{\sffamily{\thepage~\textbar\hspace{3.45cm} 1--\pageref{LastPage}}}}
\fancyhead{}
\renewcommand{\headrulewidth}{0pt} 
\renewcommand{\footrulewidth}{0pt}
\setlength{\arrayrulewidth}{1pt}
\setlength{\columnsep}{6.5mm}
\setlength\bibsep{1pt}

\makeatletter 
\newlength{\figrulesep} 
\setlength{\figrulesep}{0.5\textfloatsep} 

\newcommand{\topfigrule}{\vspace*{-1pt}%
\noindent{\color{cream}\rule[-\figrulesep]{\columnwidth}{1.5pt}} }

\newcommand{\botfigrule}{\vspace*{-2pt}%
\noindent{\color{cream}\rule[\figrulesep]{\columnwidth}{1.5pt}} }

\newcommand{\dblfigrule}{\vspace*{-1pt}%
\noindent{\color{cream}\rule[-\figrulesep]{\textwidth}{1.5pt}} }

\makeatother

\twocolumn[
  \begin{@twocolumnfalse}
{\includegraphics[height=30pt]{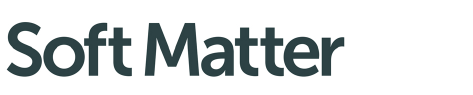}\hfill\raisebox{0pt}[0pt][0pt]{\includegraphics[height=55pt]{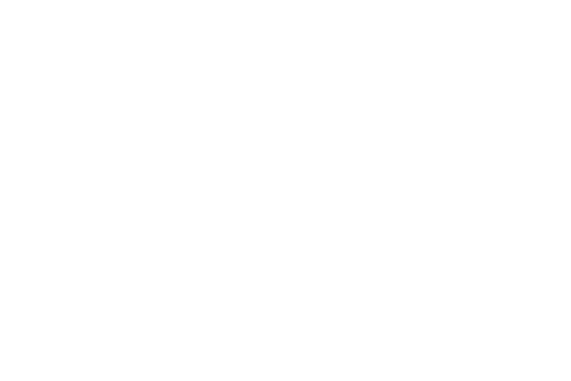}}\\[1ex]
\includegraphics[width=18.5cm]{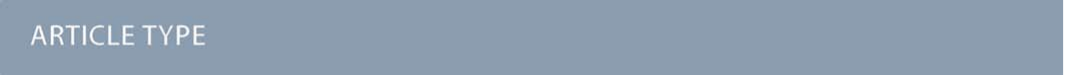}}\par
\vspace{1em}
\sffamily
\begin{tabular}{m{4.5cm} p{13.5cm} }

\includegraphics{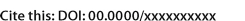} & \noindent\LARGE{\textbf{Dynamics of Phase-Separated Interfaces in Inhomogenous and Driven Mixtures$^\dag$}} \\
\vspace{0.3cm} & \vspace{0.3cm} \\

 & \noindent\large{Jacopo Romano\textit{$^{a,\ddag}$}, 
 Ramin Golestanian\textit{$^{a,b}$} and
 Beno\^it Mahault\textit{$^{a}$}}\\

\includegraphics{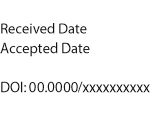} & \noindent\normalsize{We derive effective equations of motion governing the dynamics of sharp interfaces in phase-separated binary mixtures driven by spatio-temporal modulations of their material properties. We demonstrate, in particular, that spatial heterogeneities in the surface tension induce an effective capillary force that drives the motion of interfaces, even in the absence of hydrodynamics. Applying our sharp interface model to quantify the dynamics of thermophoretic droplets, we find that their deformation and transport properties are controlled by a combination of bulk and capillary forces, whose relative strength depends on droplet size. Strikingly, we show that small thermophobic droplets --composed of a material with a positive Soret coefficient-- can spontaneously migrate towards high-temperature regions as a result of capillary forces.} \\

\end{tabular}

\end{@twocolumnfalse} \vspace{0.6cm}
]

\renewcommand*\rmdefault{bch}\normalfont\upshape
\rmfamily
\section*{}
\vspace{-1cm}


\footnotetext{\textit{$^{a}$~Max Planck Institute for Dynamics and Self-Organization (MPI-DS), 37077 G\"ottingen, Germany.}}
\footnotetext{\textit{$^{b}$~Rudolf Peierls Centre for Theoretical Physics, University of Oxford, Oxford OX1 3PU, United Kingdom.}}
\footnotetext{\textit{$^{\ddag}$~Present address: SISSA—International School for Advanced Studies and INFN, via Bonomea 265, 34136 Trieste, Italy.}}

\footnotetext{\dag~Electronic Supplementary Information (ESI) available: [details of any supplementary information available should be included here]. See DOI: 10.1039/cXsm00000x/}


\section{Introduction}

Phase separation is a prototypical example of pattern formation in soft matter, and naturally emerges when the enthalpic cost of mixing multiple substances outweighs the entropic gain of maintaining a homogeneous blend. 
From a theoretical perspective, this phenomenon is routinely described by phase-field models, where the formation of patterns is captured by a conserved field whose dynamics minimizes the free energy of the system~\cite{Cahn_1958, Cahn_1959}. 

Sufficiently far from the critical point, binary phase-separated mixtures consist of nearly homogeneous domains separated by sharp interfaces. Their dynamics is then largely dominated by the motion of the interfaces, whose description involves a significantly reduced number of degrees of freedom. Deriving an effective equation governing the motion of interfaces then enables considerable simplifications in the characterization of the system's dynamics, and has therefore attracted significant interest~\cite{KawasakiPTP1982,cahn_elliott_novick-cohen_1996,ganesan1998,Shiwa1988,Elder2001,golestanian2001a,golestanian2001b,Golestanian2003,Elder2004,Barbara2015},
while similar approaches have been developed to investigate pattern formation when the order parameter is not conserved~\cite{ALLEN1979,KawasakiPTP1982_2,Bray_2002,Romano_2024}.
Notably, effective sharp interface models find a wide range of applications, including the prediction of coarsening scaling laws~\cite{Bray_2001} and of the roughening 
dynamics of fluid interfaces in porous media~\cite{KPZ,ganesan1998,Dube1999}, the description of grain boundaries~\cite{LobkovskyPRE2001} and crystal growth~\cite{Sekerka2004}, as well as the modeling of epithelium and biofilm expansion~\cite{Hallatschek2023,Wang2017}.

In many situations, phase separation occurs in the presence of external fields or spatial inhomogeneities that significantly influence the behavior of interfaces. 
Common examples include fluids subject to the influence of gravity~\cite{Shiwa1988,LacastaPRB1993} or to the presence of impurities~\cite{ganesan1998}. 
Moreover, spatio-temporal control of polymer blends or colloidal suspensions by the application of light~\cite{Voit2005PRL,Walton_2018,shin2017spatiotemporal,kim2023light} or temperature~\cite{Alt_1991, Miranville_2005, Gonnella_2008,Krekhov_2004,Rasuli2005,Rasuli2008, Anders_2012,Voit2005PRL,Voit_2007,Walton_2018,rullmann2004} gradients, or via periodic heating and cooling cycles~\cite{singh2011}, offers appealing prospects in terms of controlling the morphology of phase-separated interfaces. 

Another field where phase separation in inhomogeneous media is particularly relevant is biology. 
Biomolecular condensates, formed through liquid-liquid phase separation, play a central role in regulating key biological processes inside the cell~\cite{Banani_2017}.
In recent years, the characterization of the physics of these emulsions, and how they interact with the highly heterogeneous and dynamic cellular environment, has become the subject of an increasing amount of research~\cite{Weber_2019,zwicker2025}.

Determining how modulations of environmental properties affect the motion of interfaces is often a difficult task. The main reason is that these modulations generally induce multiple changes in the system's free energy structure, and therefore influence the dynamics in non-trivial ways. 
A straightforward example is temperature gradients, which impact all terms of the free energy, 
while the same should apply to systems driven by pH~\cite{testa2021sustained,jambon2023phase} or light~\cite{shin2017spatiotemporal,kim2023light} modulations.
To achieve a more quantitative understanding of the influence of heterogeneities on the dynamics of phase separation, having a general theory of how the spatial and temporal dependencies of all the parameters in the free energy influence the motion of interfaces is then particularly valuable.

Here, we consider a binary mixture described by a phase-field model, and derive an effective equation of motion for sharp, phase-separated interfaces in the presence of generic spatio-temporal inhomogeneities. We show that the interface dynamics is controlled by a combination of effective bulk and capillary forces, whose relative magnitude depends on the typical scale of the problem. 
In particular, our results indicate that gradients of surface tension can drive the motion of interfaces even in the absence of hydrodynamics driving Marangoni flows~\cite{marangoni1871ueber}.
We apply our framework to systems subject to temperature gradients, which enables us to characterize the thermophoretic motion of phase-separated droplets, and illustrate how their dynamics is affected by the joint influence of bulk and capillary forces. In our current treatment, we ignore the interplay between temperature inhomogeneities and fluctuations, which can manifest themselves in a variety of ways \cite{Najafi2004a,Golestanian2002a,Krueger2024}.

The paper is organized as follows: in Sections~\ref{ss:Stationary_mod_assumptions} and~\ref{sec:interface_equation}, we derive the effective sharp interface model in the presence of stationary modulations of material parameters. 
In Section~\ref{sec:1d_simple}, we then investigate a minimal example highlighting the role of capillary forces induced by gradients of the surface tension in the transport of droplets. 
In Section~\ref{Temperature_Modulation}, we apply the sharp interface theory to nonisothermal systems. Working both close to and far from criticality, we determine how the various effective forces induced by temperature gradients affect the shape and transport of droplets. We then generalize our approach to account for space- and time-dependent modulations in Section~\ref{Time_dependent}, before summarizing our results and discussing possible extensions of our framework in Section~\ref{Discussion}.

\section{Spatially modulated mixtures}
\label{Time_independent}

\subsection{Phase Separation in inhomogeneous conditions}
\label{ss:Stationary_mod_assumptions}

We consider a binary mixture of species $A$ and $B$ described by a phase field $\phi(\x,t)$, which measures the relative concentration of one of the two species (say $A$) with respect to the total concentration of $A+B$. 
This field evolves according to the continuity equation: $\partial_t\phi=-\nabla\cdot \J$ as a result of total number conservation. When the physical properties of the material vary within the sample, the mass current $\J$ is given by a straightforward generalization of model B~\cite{Halperin&Hohenberg}:
\begin{subequations}
\label{eq:fieldeq}
\begin{align}
\label{eq:Current}
    \J(\phi,\x) & = -D(\phi,\x)\nabla\mu(\phi,\x)~,
    \quad
    \mu(\phi,\x)=\dfrac{\delta \F}{\delta \phi}(\phi,\x)~,\\
\label{eq:FreeEn}
    & \F[\phi] = \int d^d x\; \left[\dfrac{K(\phi,\x)}{2}|\nabla\phi|^2+U(\phi,\x)\right]~,
\end{align}
\end{subequations}
where $d$ denotes the number of spatial dimensions.
Namely, the concentration $\phi$ is advected, up to a mobility factor $D$, by the gradient of the chemical potential $\mu$, 
which itself derives from a free energy functional $\F$.
The free energy gathers a bulk contribution, $U$, and an interface-penalizing term with an elastic constant $K$.
Note that we allow all parameters of the theory to vary both with $\phi$ and in space. 
In what follows, 
stated spatial dependencies of parameters are explicit, and thus distinct from those arising through $\phi$.

For systems with homogeneous material parameters in the phase separation regime, solutions of Eqs.~\eqref{eq:fieldeq} correspond to homogeneous domains with equilibrium bulk concentrations $\bphi_- < \bphi_+$, separated by sharp interfaces.
The thickness of these interfaces is associated with a length scale $l \simeq \sqrt{U_0/K_0}$, where $U_0$ and $K_0$ are characteristic scales extracted from $U$ and $K$, respectively. 
In addition, the relaxation dynamics of the phase field $\phi$ toward the equilibrium interface profile $\bphi$ happens over the time scale $t_l\simeq l^2/D_0$, with $D_0$ a typical mobility scale.  
The sharp interface limit is obtained when the macroscopic scales of the dynamics, such as the sizes of the phase-separated domains and the characteristic radius of curvature of their interfaces, are much larger than $l$ \cite{Bray_2002}.
This condition ensures a time scale separation between the motion of the interface and the relaxation of the concentration around its equilibrium profile, which can be considered as fast. 
A moving interface is then described at all times by the stationary concentration profile $\bphi$ which locally minimizes $\F$ and interpolates between the equilibrium concentrations $\bphi_\pm$.

The sharp interface limit naturally extends to spatially varying $K$, $U$, and $D$ in Eq.~\eqref{eq:fieldeq}, when these parameters vary over length scales that remain large with respect to $l$.
Under this assumption, the equilibrium bulk densities become local quantities that depend on the position $\x$, and can be determined from the common tangent construction~\cite{porter2009phase} applied to
the free energy density $U(\phi,\x)$.
In particular, at each point of space the chemical potential $\comtg(\x)$ and phase concentrations $\bphi_{\pm}(\x)$ are solutions of:
\begin{equation}
\label{eq:comtg}
    \begin{cases}
    \left.\partial_\phi U(\phi,\x)\right|_{\bphi_+(\x)}
    = \left.\partial_\phi U(\phi,\x)\right|_{\bphi_-(\x)}
    = \comtg(\x)~,\\
    U[\bphi_+(\x),\x]-U[\bphi_-(\x),\x] = \comtg(\x)\Dbp(\x)~,
    \end{cases}
\end{equation}
where $\Dbp(\x)=\bphi_+(\x)-\bphi_-(\x)$.
In general, spatial variations of $\comtg$ lead to nonvanishing bulk currents, due to Eq.~\eqref{eq:Current}.
To keep the corresponding mass transport slow as compared to the relaxation of $\phi$ toward the stationary interface profile, we further require variations of $\comtg(\x)$ to be smooth across the sample, i.e. $|\nabla \comtg(\x) / \comtg(\x)| \ll l^{-1}$.

In the following section, we derive an effective equation of motion for the interface separating phase-separated domains with concentrations $\bphi_{\pm}(\x)$.
This equation involves the interface surface tension $\sigma(\x)$, which is here position-dependent.
To evaluate $\sigma(\x)$, we define the bulk pressure and an adjusted free energy as
\begin{subequations}
\begin{align}
    \label{eq:Pi}
    \tilU(\phi,\x) &= U(\phi,\x)-\comtg(\x)\phi~,\\
    \label{eq:Fprime}
    \tilF[\phi] &= \F[\phi]-\int d^d x \, [\comtg(\x)\phi]~.
\end{align}
\end{subequations}
When $\comtg$ is constant, 
the stationary interface profile $\bphi$ is obtained by minimizing $\F'$, whereby $\comtg$ acts as a Lagrange multiplier ensuring mass conservation~\cite{Bray_2002}.
The surface tension, i.e. the free energy of the interface per unit area, is then obtained from Eq.~\eqref{eq:FreeEn} by evaluating $\F'[\bphi]$, as we detail in Sec.~A of ESI$\dag$ for completeness.
Analogously, for inhomogeneous model parameters, $\bphi$ is  obtained by minimizing the r.h.s.\ of Eq.~\eqref{eq:Fprime},
thus accounting for the common tangent construction~\eqref{eq:comtg} which implies that $\mu[\bphi(\x),\x] \to \comtg(\x)$ in the bulk phases.
Moreover, for smooth modulations, variations of the material parameter across the interface are negligible:  the surface tension then only depends on the local values of the coefficients of the free energy. 
As a result, its expression is a straightforward generalization of the homogeneous case, and reads
\begin{equation}
    \label{eq:surftens}
    \sigma(\x)=\int_{\bphi_-(\x)}^{\bphi_+(\x)}d\phi\,\sqrt{2 K(\phi,\x)[\tilU(\phi,\x)-\tilU_-(\x)]},
\end{equation}
where $\tilU_-(\x) \equiv \tilU[\bphi_-(\x),\x]$ is the bulk pressure in the dilute phase.

\subsection{The interface equations of motion}
\label{sec:interface_equation}

We now consider an interface in $d=3$ separating domains with concentrations $\bphi_{+}$ and $\bphi_{-}$. Denoting $G_D(\x,\y)$ as the Green's function solution of $\nabla \cdot [D(\phi,\x)\nabla G_D(\x,\y)]=\delta^3(\x-\y)$, we recast Eq.~\eqref{eq:fieldeq} as
\begin{equation}
    \label{eq:invcontinuity}
    \int d^3y \; G_D(\x,\y)\,\partial_t\phi(\y,t)=\dfrac{\delta \tilF}{\delta\phi}(\phi,\x)+\comtg(\x)-\mu_{\infty},
\end{equation}
where $\mu_{\infty}$ is an integration constant that we determine below.
In addition, for closed systems the total mass conservation imposes a second relation:
\begin{equation}
    \label{eq:conslaw}
    \int d^3 x \; \partial_t\phi(\x,t)=0.
\end{equation}
Below, we derive from Eqs.~(\ref{eq:invcontinuity},\ref{eq:conslaw}) an effective equation of motion for the interface within the sharp interface approximation.
This section outlines the main steps of the derivation, while additional details on the calculations are provided in Sec. B of ESI$\dag$.

Denoting $V_+(t)$ as the volume of the phase with density $\bphi_+$ and taking the limit $l \to 0$, we approximate the concentration field as
\begin{equation}
    \label{eq:fieldappr}
    \phi(\x,t)=\bphi_-(\x)+\Dbp(\x) \,\Theta[\x,V_+(t)]~,
\end{equation}
where $\Theta$ is the indicator function, equal to $1$ if $\x\in V_+(t)$ and to $0$ otherwise. 
The volume $V_{+}(t)$ need not correspond to a connected domain, such that the system may \textit{a priori} consist of several droplets.
The only geometrical constraint imposed by the sharp interface approximation is that the mean curvature $H$ of the surface satisfies $H l \ll 1$. 
Equation~\eqref{eq:fieldappr} then amounts to treating the interface profile as a discontinuous step between the concentrations $\bphi_{\pm}$. 

Assuming Eq.~\eqref{eq:fieldappr}, the only contribution to the time derivative of $\phi$ comes from the interface velocity $\vi$:
\begin{equation}
    \label{eq:timederappr}
    \partial_t\phi(\x,t)=-\Dbp(\x)\vi(\x,t)\cdot\nabla\Theta[\x,V_+(t)].
\end{equation}
Substituting~\eqref{eq:timederappr} into the l.h.s.\ of Eq.~\eqref{eq:invcontinuity}, 
we integrate the gradient of the step function by parts and use the divergence theorem to obtain
\begin{equation*}
    \int d^3y \; G_D(\x,\y)\,\partial_t\phi(\y,t) = \int_{\partial V_{+}(t)} d S_{\Y} \, \n(\Y)
    \cdot\vi(\Y,t) \zeta(\x,\Y)~,
\end{equation*}
where the $\Y$ integration on the r.h.s.\ is performed over the boundary of $V_+$ which defines the interface, while the normal unit vector $\n$ points outwards from $V_+$ and $\zeta(\x,\Y) \equiv \Dbp(\Y)G_D(\x,\Y)$.

Evaluating the functional derivative of $\F$ with the discontinuous profile~\eqref{eq:fieldappr} leads to a singular behaviour. 
Therefore, 
we consider an infinitesimal volume $dW$ around a point $\X$ on the interface.
The thickness of $dW$ is assumed large as compared to $l$,
such that its lateral faces orthogonal to $\n(\X)$ sit well into the bulk phases.
On the other hand, 
the dimensions of $dW$ in the two remaining directions are taken small with respect to the radius of curvature of the interface.
Following Refs.~\cite{Romano_2024, Bray_2002}, we multiply both sides of Eq.~\eqref{eq:invcontinuity} by $\n(\X)\cdot\nabla\phi$
and integrate over $dW$. 
For terms which are not singular, this procedure straightforwardly leads to a $\Delta\bphi(\X)$ prefactor,
while integrating the functional derivative of $\F'$ requires more effort. 
After some calculations detailed in ESI$\dag$, and applying a similar approach to Eq.~\eqref{eq:conslaw}, we obtain
\begin{subequations}
\label{eq:walleqs}
    \begin{align}
    \label{eq:integralmotion}
    \int_{\partial V_{+}(t)}dS_{\Y}\, \n(\Y)\cdot\vi(\Y,t) \zeta(\X,\Y) & = \nonumber\\
    \frac{1}{\Dbp(\X)} \big[2 H(\X) \sigma(\X) + & \n(\X) \cdot\nabla\sigma(\X)\big] +\comtg(\X)-\mu_{\infty}~,\\
    \label{eq:conseq}
    \int_{\partial V_+(t)}dS_{\Y}\, \n(\Y)\cdot\vi(\Y,t) \Dbp(\Y) & = 0~,
    \end{align}
\end{subequations}
where
$H(\X)$ is the mean curvature of the interface at $\X$,
while the surface tension $\sigma(\X)$ is expressed in Eq.~\eqref{eq:surftens}.
Equations~\eqref{eq:walleqs} ---and their generalization to the time-dependent case, see Eqs.~\eqref{eq:TDwalleqs}---
are the central result of this work, since when solved for $\vi(\X,t)$ they fully determine the interface dynamics.
Although they are formally written for $d = 3$, their generalization to arbitrary dimensions is straightforward, as we illustrate below\footnote{The interface equations for an arbitrary dimension $d \ge 1$ are identical to Eqs.~\eqref{eq:walleqs},  except that the term $2 H(\X) \sigma(\X)$ on the r.h.s.\ of Eq.~\eqref{eq:integralmotion} must be replaced with $(d-1) H(\X) \sigma(\X)$.}.

Before turning to concrete applications of Eqs.~\eqref{eq:walleqs}, 
we first comment on their structure and the nature of each of their terms. 
Equation~\eqref{eq:integralmotion} describes the interface motion as an overdamped dynamics, 
albeit with a nonlocal drag force consisting in the convolution of the interface velocity with the friction kernel $\zeta$. 
Such nonlocal friction originates from the conservation law on $\phi$, as can be observed from the fact that $\zeta(\X,\Y) \propto G_D(\X,\Y)$ (see Eq.~\eqref{eq:invcontinuity}).
As we show in the next sections, the nonlocality of the interface friction directly influences the scaling of a droplet mobility with its volume.

The effective drag force on the interface is balanced by the r.h.s.\ of Eq.~\eqref{eq:integralmotion}, which features three contributions.
The first one is the standard Laplace pressure~\cite{Bray_2002} which results from the elastic cost of interfaces. 
When the material properties of the system are homogeneous, this term leads to the minimization of the surface area of the domain $V_+$ while preserving its total volume. 
The second term is also a capillary force, but originates from gradients of the free energy density, and drives the interface towards regions of lower surface tension. 
Despite significant interest in droplet dynamics driven by inhomogeneous surface tension~\cite{thomson1855xlii,kamotani1995thermocapillary,karbalaei2016thermocapillarity}, this term has been, to the best of our knowledge, unnoticed so far.
Most of the existing literature instead focuses on currents induced by Marangoni flows~\cite{marangoni1871ueber,subramanian2002fluid},
which lead to qualitatively analogous transport phenomena.
This likely originates from the fact that hydrodynamic effects dominate the transport of sufficiently large droplets that move under a momentum-conserving dynamics, as can be shown from scaling arguments~\cite{Bray_2002}.
However, when flows are suppressed by the presence of a momentum sink, such as 
for systems in contact with a substrate or when phase separation happens in a porous medium~\cite{shimizu2017impact},
transport induced by the capillary terms in Eq.~\eqref{eq:integralmotion} may compete with or even dominate that arising from Marangoni flows.

The thermodynamic force $\comtg(\x)$ 
results from the fact that the current in Eq.~\eqref{eq:Current} drives the concentration $\phi$ to regions with lower free energy.
Contrary to the first two terms described above, it is a bulk contribution and thus dominates the dynamics of large droplets.
Like the gradient of the surface tension, this force vanishes when the material parameters are homogeneous, 
since a constant $\comtg$ can be absorbed in a redefinition of $\mu_{\infty}$. 
In an inhomogeneous system, 
an enlightening example of spatially-varying $\comtg(\x)$ results from the coupling of 
a homogeneous free energy $\Fdens_0(\phi)$
with an externally applied potential $V(\x)$, 
such that $\Fdens(\phi,\x)=\Fdens_0(\phi)+V(\x)\phi$. 
For instance, $V$ can be the gravitational potential leading to droplet sedimentation, such that $V(\x)=\Delta m g z$, where $\Delta m$ is the mass density difference between the two species, $g$ is the standard acceleration induced by gravity, and $z$ denotes the vertical coordinate.
Using Eqs.~\eqref{eq:comtg} and~\eqref{eq:surftens}, one concludes that $V$ does not affect the equilibrium concentrations $\bphi_\pm$ nor the surface tension, both of which remain homogeneous, while it enters the common tangent construction, such that $\comtg(\x)=V(\x)/\Dbp$.

Finally, Eq.~\eqref{eq:conseq} provides a condition allowing to determine the integration constant $\mu_\infty$.
When combined, Eqs.~\eqref{eq:walleqs} thus fully characterize the dynamics of sharp interfaces. 
In particular, the interface dynamics only depends on quantities which are directly measurable, such as the binodal densities, the surface tension, the chemical potential and the mixture mobility, and can thus be determined without solving Eq.~\eqref{eq:fieldeq} explicitly or even requiring a specific model for the free energy $\F$. 

\subsection{Dynamics driven by a surface tension profile}
\label{sec:1d_simple}

To illustrate how Eqs.~\eqref{eq:walleqs} can be applied in practice, 
we consider a minimal example of a single droplet moving in a one-dimensional inhomogeneous medium. 
In the next section, we will then focus on the case of
systems driven by inhomogeneous temperature profiles.
Here, we consider the free energy:
\begin{equation}
\label{eq:simplefreen}
    \F[\phi]=\int dx \left[ \dfrac{K}{2}(\partial_x\phi)^2+\dfrac{A(x)}{4}(\phi-\bphi_+)^2(\phi-\bphi_-)^2 \right],
\end{equation}
and assume a constant mobility $D(\phi,x)=D$. 
The free energy~\eqref{eq:simplefreen} induces modulations of the surface tension, but leaves the binodal concentrations $\bphi_\pm$ constant. 
In one dimension, a droplet of phase $+$ is described by the position of its left and right interfaces denoted $x_l$ and $x_r$, respectively.
The integral on the l.h.s.\ of Eq.~\eqref{eq:integralmotion} then becomes a point-wise evaluation at the two interfaces, and in a periodic domain of size $L$ the Green's function reads $G_D(x,y)=(2D L)^{-1}|x-y|(L - |x-y|)$ (see Sec. C of ESI$\dag$), yielding
\begin{equation}
\label{eq:interface_eq_simple}
\begin{split}
    \dfrac{\Dbp |x_r - x_l|(L - |x_r - x_l|)}{2 D L}\dot{x}_r & = -\dfrac{\sigma'(x_l)}{\Dbp} - \mu_\infty, \\
    \dfrac{\Dbp |x_r - x_l|(L - |x_r - x_l|)}{2 D L}\dot{x}_l & = -\dfrac{\sigma'(x_r)}{\Dbp} + \mu_\infty,
\end{split}
\end{equation}
where we have used that the mean curvature $H = 0$ and the chemical potential is constant. 
The surface tension is given in terms of the parameters of the theory by $\sigma(x)= \frac{1}{6\sqrt{2}}\sqrt{K A(x)}\Dbp^3$ (using Eq.~\eqref{eq:surftens}). 

The mass conservation condition~\eqref{eq:conseq} then leads to $\dot{x}_r = \dot{x}_l$, such that the droplet size $S=|x_r-x_l|$ remains constant over time. 
Making use of these relations in Eq.~\eqref{eq:interface_eq_simple}, we eliminate $\mu_\infty$ and find that the droplet centre of mass $x_{\rm c} = \tfrac{1}{2}(x_l + x_r)$ and velocity $v_{\rm c} = \dot{x}_{\rm c}$ obey
\begin{equation}
\label{eq:simpleEoM}
    v_{\rm c} = - M_{\rm eff} U_{\rm eff}'(x_{\rm c})~,
\end{equation}
where the effective mobility and potential read 
\begin{equation*}
    M_{\rm eff} = \frac{D L}{(\Dbp)^2 S(L-S)}, \quad
    U_{\rm eff}(x_{\rm c}) = \sigma\left(x_{\rm c} + \frac{S}{2}\right)+\sigma\left(x_{\rm c} - \frac{S}{2}\right)~.
\end{equation*}
Equation~\eqref{eq:simpleEoM} highlights an important feature of the dynamics: in the presence of smooth changes in the material parameters, a droplet moves so as to minimize the total surface tension at its interfaces under the constraint of mass conservation. 

\begin{figure}[t!]
    \centering
    \includegraphics[width=\linewidth]{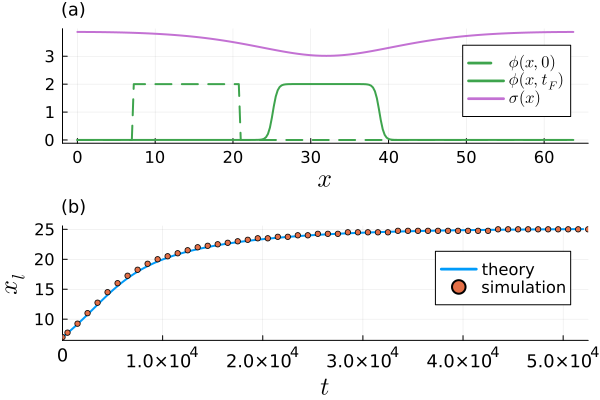}
    \caption{Droplet dynamics driven by the free energy density~\eqref{eq:simplefreen} in a one-dimensional periodic domain.
    (a) As a result of the inhomogeneous surface tension profile (pink line), a droplet at $t=0$ (green dashed line) migrates 
    toward the centre of the domain,
    where it stabilizes after a sufficiently long time $t_F$ (full green line). 
    (b) Position of the left interface of the droplet vs.\ time as predicted by the sharp interface approximation (Eq.~\eqref{eq:simpleEoM}), 
    and measured in numerical simulations. 
    Additional details on the simulations are given in Sec. D of ESI$\dag$.
    }
    \label{fig:examplederiv}
\end{figure}

To verify that Eq.~\eqref{eq:simpleEoM} faithfully describes the dynamics, we confront it with numerical simulations of the field equations~\eqref{eq:fieldeq} using the free energy \eqref{eq:simplefreen}.
In particular, we impose a sinusoidal modulation of $A(x)$, which exhibits a unique minimum at $x = L/2$
(See Sec. D of ESI$\dag$ for details about numerical simulations and the values of the parameters used).
As noted previously, spatial variations of $A$ induce an inhomogeneous surface tension profile, which is also minimal at the centre of the domain (Fig.~\ref{fig:examplederiv}(a)).
Hence, droplets spontaneously migrate towards this position.
Note that since the effective force in Eq.~\eqref{eq:simpleEoM} is evaluated at the droplet interfaces, 
for asymmetric profiles the stationary position of its centre of mass does not necessarily coincide with the location of minimum surface tension.
Figure~\ref{fig:examplederiv}(b) further shows that, beyond this qualitative picture,
Eq.~\eqref{eq:simplefreen} quantitatively captures the droplet trajectory, as reflected by the excellent agreement between the theoretical curve and the data points obtained from simulations.

The above analysis can be easily generalized to dimensions $d = 2$ and 3.
Considering a spherical droplet of radius $R$ with constant $\bphi_\pm$ and $\comtg$ in a surface tension profile $\sigma(\x)$,
its shape remains unaffected by smooth modulations of $\sigma(\x)$.
Using Eqs.~\eqref{eq:walleqs}, we then derive the equation governing the evolution of its centre of mass $\x_{\rm c}$ in Sec. E of ESI$\dag$.
Working in an infinite space for simplicity, we find that the droplet velocity $\vi_{\rm c} = \dot{\x}_{\rm c}$ obeys 
$\vi_{\rm c} = -M_{\rm eff}\nabla U_{\rm eff}(\x_{\rm c})$, with
\begin{align} \label{eq:mobandpotd_dim}
M_{\rm eff} = \frac{d D}{(\Dbp)^2 R^d V_d}, \qquad
U_{\rm eff}(\x_{\rm c})= \int_{S_R(\x_{\rm c})} dS_{\Y} \,\sigma(\Y).
\end{align}
Equation~\eqref{eq:mobandpotd_dim} generalizes the above expression to higher dimensions in the limit $L \to \infty$. The effective potential $U_{\rm eff}$ is now obtained by integrating the surface tension over the interface,
such that $S_R(\x_{\rm c})$ denotes the sphere of radius $R$ centred in $\x_{\rm c}$, while $V_d$ is the volume of the unit $d$-sphere. 
Contrary to the motion induced by Marangoni flows for which the mobility scales as $R^{-1}$ as a result of Stoke's drag~\cite{schmitt2016marangoni}, the mobility in Eq.~\eqref{eq:mobandpotd_dim} is inversely proportional to the droplet volume.
Hence, the velocity deriving from Eq.~\eqref{eq:mobandpotd_dim} should dominate the dynamics of small droplets, while hydrodynamic effects become dominant only when $R$ is sufficiently large.

\section{Temperature Modulations}

\label{Temperature_Modulation}

\subsection{Nonequilibrium thermodynamics of phase separation}

The physical properties of a mixture can be straightforwardly modulated by applying an inhomogeneous temperature profile across the sample.
The study of nonisothermal phase separation has received a great deal of attention both from the theoretical~\cite{Alt_1991, Miranville_2005, Gonnella_2008,Krekhov_2004, Anders_2012} and experimental~\cite{Okinaka1995,Voit2005PRL,Voit_2007,Walton_2018} viewpoints.
Polymer blends, in which precise temperature modulations can be obtained by focusing laser beams inside the sample~\cite{Voit2005PRL,Voit_2007}, are particularly well-suited to explore this phenomenon.

The dynamics of a mixture in an inhomogeneous temperature profile $T(\x)$ can be modelled by means of nonequilibrium thermodynamics~\cite{Mazur_2013}.
The starting point is a generalization of the current in Eq.~\eqref{eq:fieldeq}, which reads
\begin{equation}
\label{eq:tempCH}
    \J[\phi,T(\x)] = - D(\phi,T)\nabla\mu(\phi,T)-D_T(\phi,T)\nabla T,
\end{equation} 
where we suppose that $\J$ varies in space only through $\phi$ and $T$
The dependencies of the mobilities $D$ and $D_T$ on the chemical concentration $\phi$ and temperature are generally nontrivial, and various models have been considered in the literature~\cite{Anders_2012,cahn_elliott_novick-cohen_1996}.
Here, we use the forms
\begin{align}
\label{eq:diffusivitiesDS}
    D(\phi) = D_0\,\phi(1-\phi), \qquad
    D_T(\phi) = D_{T,0}\,\phi(1-\phi),
\end{align}
which ensure that the current~\eqref{eq:tempCH} vanishes when only a single species is present ($\phi=0$ or $\phi=1$). 
For simplicity, we have also neglected the dependencies of the mobilities in the temperature, 
as for sufficiently smooth modulations they do not significantly affect the dynamics of droplets.

The current in Eq.~\eqref{eq:tempCH} differs from the expression~\eqref{eq:Current} as it includes an additional term modeling thermodiffusion (or Soret effect). 
Although generalizing Eqs.~\eqref{eq:walleqs} by taking into account this term is possible, 
we show below that many physically relevant scenarios admit an expression for $\J$ analogous to Eq.~\eqref{eq:Current}, where the contribution from the Soret current enters as an additional term in the free energy $\F$.
In what follows, and for the sake of simplicity, we restrict the analysis to these cases. 

We model the bulk potential of the free energy density with the Flory-Huggins theory~\cite{Flory_1941,Huggins_1942}:
\begin{align}
    \label{eq:FH}
    U(\phi,T)=k_B T\bigg[\frac{\phi}{N_A}\log(\phi)+\frac{1-\phi}{N_B}\log(1-\phi)
    +\chi\phi(1-\phi)\bigg],
\end{align}
where  $N_A$ and $N_B$ are degrees of polymerization of species $A$ and $B$, respectively. 
The coefficient $\chi$  is the interaction parameter, 
which can be expressed as a function of the temperature as~\cite{Wignall85,SchmidBook}
\begin{equation}
    \label{eq:chitempdep}
    \chi=\alpha+\frac{\beta}{T},
\end{equation}
where $\alpha$ and $\beta$ are phenomenological parameters that can be fitted from experimental data.
Additionally, we model the elastic contribution to the free energy using de Gennes' random phase approximation \cite{deGennes1980,Krekhov_2004}:
\begin{equation}
    \label{eq:RPAelastic}
    K(\phi,T)=\dfrac{k_B T}{18}\left[\dfrac{\theta_A^2}{\phi}+\dfrac{\theta_B^2}{1-\phi}\right]~,
\end{equation}
where $\theta_{A,B}$  are the Kuhn lengths associated with $A$ and $B$, respectively.
In the presence of an inhomogeneous temperature profile, the chemical potential for the binary mixture is then given by~\cite{Gonnella_2008}:
\begin{equation}
    \label{eq:nonisomu}
    \mu=\dfrac{\delta}{\delta\phi}\left\{\int \dfrac{d^d x}{ k_B T(\x)}\left[\dfrac{K[\phi,T(\x)]}{2}|\nabla\phi|^2+U[\phi,T(\x)]\right]\right\}.
\end{equation}
With the bulk potential~\eqref{eq:FH}, 
the system exhibits phase separation when the interaction parameter $\chi$ overcomes the critical value $\chi_c=(\sqrt{N_A}+\sqrt{N_B})^2/(2 N_A\,N_B)$,
which in turn determines the critical temperature $T_c$ through Eq.~\eqref{eq:chitempdep}.
In the remainder of this section, 
we analyze the dynamics of droplets in systems maintained close to the critical point ($T(\x) \lesssim T_c$) and far from it ($T(\x) \ll T_c$).

\subsection{Dynamics close to the critical point}
\label{subsec:closetocritical}

We focus here on two-dimensional mixtures maintained close to their critical point. 
In particular, we show below that Eqs.~\eqref{eq:walleqs} can be used to quantify the shape deformations of a droplet in response to an anisotropic temperature profile,
whose amplitude can be directly related to the Soret coefficient of the mixture. 
Following Refs.~\cite{enge2004thermal,meier1992critical}, 
we further provide numerical estimates of the parameters of the theory for a $55\%$/$45\%$ polydimethylsiloxane/polyethylmethylsiloxane (PDMS/PEMS) mixture (see Table~\ref{tbl:coeffs}), 
and use these values to guide the approximations made in the calculation.

As $T$ is kept close to $T_c$,  
the elastic coefficient and diffusivities can be approximated by their values at the critical point: $\bK \equiv K(\phi_c,T_c)$, $\bD \equiv D(\phi_c)$, and $\bDT \equiv D_T(\phi_c)$, where $\phi_c=1/(1+\sqrt{N_A/N_B})$.
We also simplify the bulk potential~\eqref{eq:FH} by expanding it up to fourth order in $\varphi \equiv \phi-\phi_c$, and to linear order in $\tau(\x) \equiv (T(\x)-T_c)/T_c$. 
The dynamics of the concentration then obeys $\partial_t\varphi= \bD \nabla^2\delta \F_{\rm{GL}}/\delta \varphi$ with the effective Ginzburg-Landau free energy
\begin{equation}
\label{eq:EffectiveGL}
    \F_{\rm{GL}}[\varphi]=\int d^2 x\; \left\{\frac{\bK |\nabla\varphi|^2}{2k_B T_c} 
    + \left[a\;\tau(\x)\varphi+\frac{b}{2}\tau(\x )\varphi^2+\frac{c}{4}\varphi^4\right] \right\},
\end{equation}
while the coefficients resulting from the expansion are given by
\begin{align}
    \label{eq:effectiveGLcoeffs}
    a & = \frac{\bD_T T_c}{\bD}-\frac{\beta(\sqrt{N_A}-\sqrt{N_B})}{T_c(\sqrt{N_A}+\sqrt{N_B})}, \quad
    b = \frac{2\beta}{T_c}, \quad
    c=\frac{4\chi_c^2}{3}\sqrt{N_A N_B}.
\end{align}
Note that the chemical potential~\eqref{eq:nonisomu} is adimensional, and that thermodiffusion only enters through the $\bar{D}_T$-dependency of the coefficient of its linear term,
while temperature modulations also affect its quadratic term, which explicitly depends on $\tau(\x)$.
The spatially-dependent surface tension, binodal densities, and bulk chemical potential then take the compact forms: 
\begin{equation} \label{eq:surface-tension-CP}
\sigma(\x)
=\frac{2}{3c}\sqrt{\frac{2 \bK b^3}{k_B T_c}}|\tau(\x)|^{3/2},\quad
\Dbp(\x)=\sqrt{\frac{b}{c}|\tau(\x)|}, \quad
\comtg(\x)=a \tau(\x).
\end{equation}

\begin{table}[t!]
\centering
\small
  \caption{Coefficients entering the effective free energy~\eqref{eq:EffectiveGL} and the mixture mobility.
  The numerical values correspond to a PDMS/PEMS blend and are based on estimates taken from Refs.~\cite{enge2004thermal,meier1992critical}.}
  \label{tbl:coeffs}
  \def\arraystretch{1.2}
  \begin{tabular*}{0.4\textwidth}{@{\extracolsep{\fill}}lcc}
    \hline
    \;\;Coefficient & Value & Unit \\
    \hline 
     \;\; $a$ & 0.21 & \\
     \;\; $b$ & $2.06\times10^{-2}$& \\
     \;\; $c$ & $2.25\times10^{-2}$ & \\
     \;\;$\bK/(k_B T_c)$ & $8.4\times10^{-2}$ & ${}\unit{nm^2}$\\
     \;\;$\bD$ & $75.1$& $\unit{\mu m^2 s^{-1}}$\\
     \;\;$T_c$ & $ 313 $ & $ \unit{K} $ \\
     \;\;$N_A$ & $ 219.4 $ & $  $ \\
     \;\;$N_B$ & $ 257.25 $ & $  $ \\
    \hline
  \end{tabular*}
\end{table}

As a first simplification, we note that for the PDMS/PEMS mixture $N_A \simeq N_B$ (see Table~\ref{tbl:coeffs}),
such that the coefficient $a$ entering the expression of $\comtg$ is dominated by the Soret coefficient $\ST \equiv \bD_T / \bD$.
Additionally, we write the reduced temperature $\tau(\x)$ as the sum of a constant and a spatially-modulated part: $\tau(\x)=\Delta T/T_c+\delta T/T_c\; m(\x)$, where $m(\x)$ is a smoothly varying function of order one. 
In the phase-separated regime, $\Delta T<0$, 
while we assume $|\delta T| \ll |\Delta T|$.
This approximation is also generally well-verified experimentally~\cite{Voit2005PRL}, as exemplified by Ref.~\cite{enge2004thermal,meier1992critical} which reports $\Delta T=-2.5\unit{K}$ and $\delta T\simeq0.2\unit{mK}$. 

We now consider a circular phase-separated droplet of the phase $+$ in two dimensions, and denote its radius as $R_0$.
We further assume, without loss of generality, that the mobility $D_T$ is negative, since the opposite case $D_T > 0$ is equivalent upon considering a droplet of the complementary phase. 
Heating a localized spot of the sample, e.g. by means of a laser~\cite{Voit_2007}, the droplet quickly migrates by thermophoresis towards its centre, and settles at some fixed position $\x = 0$.
Assuming a smoothly-varying temperature profile, we expand the modulation as $m(\x) \simeq 1 + \tfrac{1}{2}[m_x\left(x/R_0\right)^2+ m_y\left(y/R_0\right)^2]$ where $m_x$ and $m_y$ are both negative.
For $m_x\neq m_y$ the droplet deforms as a result of the anisotropy of the heat source, which can be quantified in the stationary regime by balancing the effective force terms in the r.h.s.\ of Eq.~\eqref{eq:integralmotion}.

While the Laplace pressure maintains the circular shape of the droplet, both surface tension gradients and the Soret current are responsible for its deformation.
To evaluate their relative magnitudes, we use Eq.~\eqref{eq:surface-tension-CP}, which gives
\begin{equation}
    \left|\frac{\nabla\sigma(\x)}{\Delta\bphi(\x)\comtg(\x)}\right| 
    \simeq
    \sqrt{\frac{\bK}{k_B T_c}\frac{b^2}{c a^2}}\left|\frac{\delta T\nabla m(\x)}{\Delta T}\right|~.
\end{equation}
Using $|\nabla m(\x)| \simeq R_0^{-1}$ at the interface, as well as the parameters listed in Table~\ref{tbl:coeffs} and above,
we find that the force resulting from gradients of $\sigma$ is completely negligible when $R_0$ is larger than a few nanometers, such that we can safely ignore it.
In addition, expanding $\sigma(\x)$ and $\Delta\bphi(\x)$ in powers of $\delta T$, we also find that the leading order corrections to these quantities induced by modulations of the temperature are negligible when considering the parameters in Table~\ref{tbl:coeffs}. 

Denoting $R(\theta)=R_0+\delta R(\theta)$ the radius of the deformed droplet as a function of the polar angle $\theta$, 
Eq.~\eqref{eq:integralmotion} in the stationary state then yields
\begin{equation}
    \label{eq:stat}
    \dfrac{\sigma^*}{\Dbp^*}H(\theta)+\ST T_c  \tau[\bgamma(\theta)]-\mu_{\infty}=0,
\end{equation}
where $\sigma^*$ and $\Dbp^*$ are obtained from Eq.~\eqref{eq:surface-tension-CP} by approximating $\tau \simeq \Delta T/T_c$, 
$\bgamma(\theta)=[R_0+\delta R(\theta)](\cos\theta,\sin\theta)$ parametrizes the interface in polar coordinates,
while we have used that $a \simeq \ST T_c$ for $N_A \approx N_B$.

Using the explicit expression for the interface curvature in polar coordinates: $H(\theta) = |R^2 + 2 R'^2 - R R''|/(R^2 + R'^2)^{3/2}$ (where $\theta$-dependencies are kept implicit),
we expand Eq.~\eqref{eq:stat} to linear order in $\delta R$, $\delta T_x$, $\delta T_y$, $\delta \mu_{\infty}$, and obtain
\begin{align}
    \label{eq:expandcurv}
    \dfrac{\sigma^*}{\Dbp ^*}\frac{\delta R''+\delta R}{R_0^2} - \frac{\delta T (m_x-m_y)}{2}\left(\ST\cos2\theta+\frac{1}{T_c}\right) + \delta\mu_{\infty}=0~.
\end{align}
Once again, we eliminate the constant $\delta \mu_{\infty}$ by imposing mass conservation, which here amounts to $\int_0^{2\pi}d\theta\delta R(\theta)=0$. 
Integrating Eq.~\eqref{eq:expandcurv} over $\theta$ and discarding vanishing terms, we then find that $\delta \mu_{\infty}=\tfrac{1}{2}\delta T(m_x-m_y)/T_c$.
Replacing this expression into Eq.~\eqref{eq:expandcurv} and solving for $\delta R$, the relative deformation induced by the inhomogeneous temperature profile reads
\begin{equation}
    \label{eq:deltaR}
    \frac{\delta R(\theta)}{R_0}= -\frac{\ST}{6}\frac{\Dbp^* R_0}{\sigma^*}\delta T(m_x-m_y)\cos2\theta~.
\end{equation}

\begin{figure}[t!]
    \centering
    \includegraphics[width=0.65\linewidth]{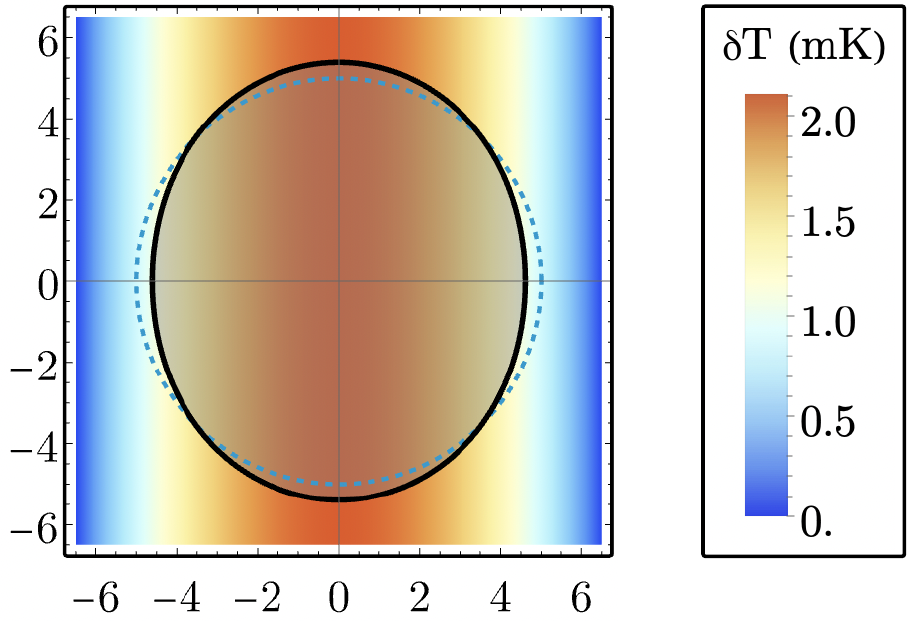}
    \caption{Deformation of a droplet trapped at the centre of a temperature stripe as predicted by Eq.~\eqref{eq:deltaR} (continuous black line). The radius of the undeformed droplet is $R_0=5\unit{\mu m}$, as marked by the dashed circle, while temperature modulations are parametrized by $\delta T = 0.1 \,\unit{mK}$, $m_x = -1$, $m_y=0$. Other parameters take the values listed in Table~\ref{tbl:coeffs}, while the color map shows local variations of the temperature.}
    \label{fig:defdroplet}
\end{figure}

Equation~\eqref{eq:deltaR} highlights the competition between the Soret effect driving to the deformation of the droplet in the inhomogeneous temperature profile, and capillary forces working to restore its circular shape. 
Experimentally, Soret-induced droplet deformations could be observed by heating a localized stripe in the sample ($m_y = 0$), whereby the droplet would acquire an elliptical shape whose major axis is aligned with the stripe, as pictured in Fig.~\ref{fig:defdroplet}.
Denoting $r=R(\pi/2)/R(0)$ as the ratio between the ellipse's minor and major axes, we then find
\begin{equation}
    \label{eq:Diffratio}
   \ST=\frac{6\sigma^*}{\Dbp^* R_0}
   \frac{1}{\delta T m_x}
   \frac{r-1}{r+1}~.
\end{equation}
Interestingly, Eq.~\eqref{eq:Diffratio} enables the measurement of the Soret coefficient of the mixture $\ST = \bD_T / \bD$ in the phase separated regime directly from the deformations of the stationary droplet shape.
Furthermore, Eq.~\eqref{eq:Diffratio} only involves parameters such as the surface tension $\sigma$ and binodal concentrations $\bphi_\pm$, which can be accessed directly in experiments.
Hence, it can be used without assuming a specific model for the free energy $\F$ or evaluate its coefficients individually.

To end this section, we briefly comment on the typical droplet size for which deformations induced by Soret currents become appreciable.
Combining the values of Table~\ref{tbl:coeffs} and Eq.~\eqref{eq:surface-tension-CP}, we estimate $\sigma^*/\Dbp^* \approx 1.5*10^{-4}\unit{nm}$ and $\ST T_c \simeq a \approx 0.21$, while following Ref.~\cite{enge2004thermal,meier1992critical} we use $\delta T / T_c \approx 10^{-6}$.
Setting $\cos2\theta = 1$ in the r.h.s.\ of Eq.~\eqref{eq:deltaR},
we thus deduce that a relative droplet deformation of a few percent is achieved as soon as $R_0 \gtrsim 10\unit{n m}$. 
In turn, a droplet of typical radius $R_0 \approx 100\unit{\mu m}$ is significantly deformed by a surprisingly weak relative temperature modulations of the order of $\delta T/T_c \gtrsim 10^{-8}$.  
We note, however, that the parameters reported in Table~\ref{tbl:coeffs} could underestimate the surface tension and/or overestimate the magnitude of the Soret coefficient.
While a discussion on the validity of these parameters in the phase coexistence regime is beyond the scope of this work, 
we note that Eqs.~(\ref{eq:deltaR},\ref{eq:Diffratio}) provide a direct method to measure the relative strengths of the effective forces induced by the Laplace pressure and the Soret currents. 

\subsection{Dynamics far from the critical point}

We now assume that the system is far from criticality, which is achieved for $T \ll T_c$. 
In this regime, the mixture is almost completely phase-separated, so that $\bphi_-\to 0$ and $\bphi_+ \to 1$.
Since the resulting concentration field varies significantly when crossing an interface, the $\phi$-dependency of the mobilities appearing in Eq.~\eqref{eq:diffusivitiesDS} cannot be neglected. 
On the other hand, we keep ignoring the variations of $D_0$ and $D_{T,0}$ with the temperature, as we are interested in the effect of weak modulations.
Under this assumption, the Soret current in Eq.~\eqref{eq:tempCH} can again be treated as an additional contribution to the chemical potential~\eqref{eq:nonisomu} which renormalizes the bulk free energy as $U[\phi,T(\x)]/[k_BT(\x)] \to U[\phi,T(\x)]/[k_BT(\x)] + \Sigma_{T,0} \phi T(\x)$, where $\Sigma_{T,0} = D_{T,0}/D_0$.
Hence, we can directly deduce the dynamics of droplets from Eqs.~\eqref{eq:walleqs}.

We demonstrate below that, far from the critical point, the capillary term in Eq.~\eqref{eq:integralmotion} arising from gradients of the surface tension is not necessarily negligible in comparison to the Soret effect. 
This contribution, in particular, dominates the thermophoretic transport of small droplets, and may either amplify or compete with the Soret effect.

Using Eqs.~\eqref{eq:comtg} and~\eqref{eq:FH} we express the binodal densities in the vanishing temperature regime as
\begin{align} \label{eq:solcomtg}
    \bphi_-(\x) \simeq 
    e^{N_A/N_B - 1 - N_A\chi(\x)}
    , \qquad 
    \bphi_+(\x) \simeq 1- e^{N_B/N_A - 1-N_B\chi(\x)},
\end{align}
while $\comtg(\x) = \Sigma_{T,0} T(\x)$ and the interaction coefficient $\chi(\x)$ is determined by the local temperature through Eq.~\eqref{eq:chitempdep}.
The binodal densities $\bphi_+$ and $\bphi_-$ converge exponentially fast to 0 and 1, respectively, as $T \to 0$ (or $\chi \to \infty$), such that we safely assume $\Dbp(\x) \simeq 1$ below.
The approximation~\eqref{eq:solcomtg}, on the other hand, is necessary to evaluate the mobilities in both phases from Eq.~\eqref{eq:diffusivitiesDS}.
To evaluate the surface tension,
we note that for large $\chi$ the entropic contribution to $U$ in Eq.~\eqref{eq:FH} is subdominant.
The bulk pressure then reads $\Pi(\phi,\x) \simeq \chi(\x)\phi(1-\phi)$,
which we combine with $K[\phi,T(\x)]$ in Eq.~\eqref{eq:RPAelastic} to calculate the integral in Eq.~\eqref{eq:surftens}. Using $\bphi_+\simeq1$ and $\bphi_-\simeq0$, we obtain
\begin{equation}
    \label{eq:surftensDS}
    \sigma(\x) \underset{\chi \to \infty}{\simeq}
    \frac{2}{9}\left(\theta_B+\dfrac{\theta_A^2}{\theta_A+\theta_B}\right)\sqrt{\chi(\x)}
    \equiv \sigma_0\sqrt{\chi(\x)}~.
\end{equation}

In what follows, we consider a droplet of concentration $\bphi_+$ in a background medium of concentration $\bphi_-$. 
For simplicity, we restrict the calculations to one-dimensional systems, as we do not expect higher-dimensional dynamics to introduce qualitative changes.
As before, we assume a weak temperature modulation $\delta T(\x)$ around a homogeneous value $T_0$, with $|\delta T(\x)| \ll T_0$.
This inhomogeneous temperature profile translates into a modulation of the Flory-Huggins interaction parameter
\begin{equation} \label{eq:chi_appmod}
    \chi(\x) \simeq \chi_0 + \delta\chi(\x),
\end{equation}
with $\chi_0 = \alpha + \beta/T_0$ and $\delta\chi(\x) = -\beta \delta T(\x) / T_0^2$ at linear order, as can be deduced from Eq.~\eqref{eq:chitempdep}.

To evaluate the l.h.s.\ of the droplet equation of motion~\eqref{eq:integralmotion}, we need to calculate the Green's function $G_D(x,y)$ solution of the equation  $\partial_x [ D(\phi) G_D(x,y)]=\delta(x-y)$.
We do this by assuming that the concentration field around the droplet is well-described by the sharp interface approximation, Eq.~\eqref{eq:fieldappr},
such that the equation for $G_D(x,y)$ is solved piecewise.
The mobilities inside and outside the droplet are given by
\begin{align} \label{eq:mob_far_from_Tc}
    D(\phi) = D_0 \phi(1-\phi) \simeq \begin{cases}
        D_0(1 - \bphi_{+,0}) \equiv \Din & \text{inside} \\
        D_0 \bphi_{-,0} \equiv \Dout & \text{outside}
    \end{cases},
\end{align}
where we have approximated the binodal densities $\bphi_{\pm}$ with their background values obtained from Eq.~\eqref{eq:solcomtg} by replacing $\chi(\x)$ with $\chi_0$, 
while we have used that for large $\chi_0$ (low $T_0$) $\bphi_{+,0} \approx 1 - \bphi_{-,0} \approx 1$.
After some calculations, we obtain
\begin{align}
\label{eq:unevenGF}
    & G_D(x,y) = 
    \frac{|x-y|}{2} \left\{ \Dout^{-1}+
    \left(\Din^{-1}-\Dout^{-1}\right)\left[\Heav(x-x_l)\Heav(x_r-x)\right] \right\} \nonumber \\
    & + \left(\Din^{-1}-\Dout^{-1}\right) \left[\frac{|x_r-y|}{2}\Heav(x-x_r)-\frac{|x_l-y|}{2}\Heav(x-x_l)\right],
\end{align}
where $\Heav(x)$ is the Heaviside step function,
while $x_l$ and $x_r$ denote the positions of the left and right interfaces, respectively.

Using Eqs.~(\ref{eq:solcomtg},\ref{eq:surftensDS},\ref{eq:chi_appmod}\ref{eq:unevenGF}), we are now able to evaluate all terms of Eqs.~\eqref{eq:walleqs} explicitly.
Following a calculation similar to that detailed in Sec.~\ref{sec:1d_simple}, 
we find that the velocity of the droplet obeys 
\begin{align}
\label{eq:dropvelDS}
    & \frac{v}{M_{\rm eff}} = \frac{\sigma_0\beta}{\sqrt{\chi_0} T_0}
    \frac{\delta T'(x_r) + \delta T'(x_l)}{2T_0}
    -\Sigma_{T,0}\left[\delta T(x_r) - \delta T(x_l)\right], 
\end{align}
where $M_{\rm eff} = \Din/S$ with $S=|x_r-x_l|$ the size of the droplet, while we have used $\Dbp \approx 1$.
As before, the velocities of the left and right interfaces are identical due to mass conservation, such that $S$ remains constant over time.

Examining the r.h.s.\ of Eq.~\eqref{eq:dropvelDS}, 
we immediately note that inhomogeneous temperature profiles induce two different effective forces driving the droplet motion.
The Soret current leads to a term that depends on the temperature difference between the two interfaces.
The resulting force, therefore, drives droplet up or down temperature gradients whenever $\Sigma_{T,0} < 0$ or $\Sigma_{T,0} > 0$, respectively.
On the other hand, the capillary term induced by gradients of the surface tension scales with the derivatives of the temperature modulation.
As $\beta > 0$~\cite{SchmidBook}, this term results in a thermophilic motion of the droplet, which seeks to reach high temperature regions where the surface tension is lower. 

In mixtures presenting a positive Soret coefficient, the two forces on the r.h.s.\ of Eq.~\eqref{eq:dropvelDS} therefore compete to determine the net direction of motion.
Since Soret currents emerge from the bulk while the capillary force originates from interfaces, their relative magnitude is controlled by the droplet size.
To illustrate this, we expand the smoothly varying temperature profile around the droplet as $\delta T(x) \simeq \Texp x$. 
Assuming $\Sigma_{T,0} > 0$ and compensating the two terms on the r.h.s.\ of Eq.~\eqref{eq:dropvelDS}, 
we find that the droplet velocity changes sign when its size reaches the critical value
\begin{equation}
    \label{eq:critical_size}
    S_c = \frac{1}{\Sigma_{T,0}T_0}
    \frac{\sigma_0\beta}{T_0\sqrt{\chi_0}}~.
\end{equation}
Our analysis thus predicts that small droplets with $S < S_c$ are always thermophilic, while large ones with $S > S_c$ become thermophobic. 
Although Eq.~\eqref{eq:critical_size} was derived assuming a one-dimensional system, we expect similar relations to hold in higher dimensions.
In particular, the fact that the motion of large droplets is dominated by the bulk Soret currents, while small droplets primarily move as a result of capillary forces, should remain independent of the dimension.

We conclude this section by commenting on nonlinear effects influencing the droplet mobility.
For simplicity, we first consider the case of an isolated droplet evolving in an infinite background of the complementary phase.
The effective mobility appearing in Eq.~\eqref{eq:dropvelDS} for a droplet of phase ``$+$'' (in which species $A$ is most abundant) is set by the ratio $\Din/S$.
From Eq.~\eqref{eq:mob_far_from_Tc}, $\Din$ is governed by the binodal density $\bphi_{+,0}$, and vanishes as the temperature approaches zero. 
Conversely, a droplet of phase ``$-$'' (in which species $B$ is most abundant) moves with a mobility determined by $\Dout$, which is proportional to $\bphi_{-,0}$ and also vanishes at zero temperature. 
Interestingly, for large $\chi_0$ their ratio calculated using Eqs.~\eqref{eq:diffusivitiesDS} and~\eqref{eq:solcomtg},
\begin{equation}
    \label{eq:diffratioDS}
    \dfrac{\Din}{\Dout} \simeq \exp\left[\frac{N_B^2-N_A^2}{N_AN_B}+(N_A-N_B)\chi_0\right],
\end{equation}
becomes highly sensitive to the difference $N_A-N_B$ between the species degrees of polymerization. 
Specifically, assuming one of the species to be much smaller than the other, the dynamics of droplets in which it is dominant is strongly suppressed as compared to droplets of the other phase.

The situation is more subtle in finite systems, where the effective droplet mobility varies with the type of imposed boundary condition.
For periodic systems, in particular, $M_{\rm eff}$ can \textit{a priori} depend on both $D_+$ and $D_-$. 
However, taking $\chi_0 \to \infty$ 
the ratio~\eqref{eq:diffratioDS} vanishes (diverges) exponentially fast when $N_B > N_A$ ($N_A > N_B$), such that the mobility dependency in $D_+$ ($D_-$) should become negligible.
Hence, in this case,
$M_{\rm eff}$ will be dominated by the value of the field mobility associated with the phase in which the largest species is most abundant.


\section{Dynamics under spatio-temporal modulations}
\label{Time_dependent}

In this section, we extend the above formalism to account for situations where the material properties of the sample are modulated both in space and time.
As before, we assume that the coefficients in the free energy vary on a characteristic length scale much larger than the interface length $l$, 
which ensures the validity of the sharp interface approximation.
In addition, we require that they vary over time scales much larger than $t_l \simeq l^2/D$, 
the typical time associated with the relaxation of the concentration field toward the stationary interface profile.

Using the time and length scale separation between the interface profile dynamics and the applied modulations, the surface tension, binodal densities, and bulk chemical potential can then be deduced from Eqs.~(\ref{eq:comtg},\ref{eq:surftens}).
Note, however, that they now all vary both in space and time, such that we generalize Eq.~\eqref{eq:fieldappr} as
\begin{equation}
    \label{eq:TDfieldappr}
    \phi(\x,t)=\bphi_-(\x,t)+\Dbp(\x,t) \,\Theta[\x,V_+(t)]~.
\end{equation}
Plugging this ansatz in Eqs.~\eqref{eq:invcontinuity} and~\eqref{eq:conslaw}, we derive the effective equations of motion for the interface
in a similar manner as presented in Sec.~\ref{sec:interface_equation} for stationary modulations.
The main difference induced by the time dependency of the binodal densities is the emergence of a new effective force driving the motion of the interface, which takes the form
\begin{equation}
\label{eq:dispforce}
    \Xi(\X,t)= -\sum_{k=\pm}\int_{V_k(t)} d^3 y\; G_D(\X,\y,t) \, \partial_t\bphi_k(\y,t),
\end{equation}
where, as before, $\X$ denotes an arbitrary point on the interface while the Green's function $G_D(\x,\y,t)$ is the same as that appearing in Eq.~\eqref{eq:invcontinuity}.
This additional term results from the fact that spatio-temporal modulations of the binodal concentrations naturally induce spatio-temporal variations of the bulk concentrations.
The resulting mass transport across the system then acts as an effective force on the interface.

Taking this new force into account, we find that the interface velocity $\vi$ now obeys
\begin{subequations}
\label{eq:TDwalleqs}
    \begin{align}
    \label{eq:TDintegralmotion}
    & \int_{\partial V_{+}(t)}dS_{\Y} \,
    \n(\Y)\cdot\vi(\Y,t) \,\zeta(\X,\Y,t) = \Xi(\X,t) + \comtg(\X,t)-\mu_{\infty} \nonumber\\
    & \qquad\qquad + \frac{1}{\Dbp(\X,t)} \big[2H(\X)\sigma(\X,t)+ \n(\X)\cdot\nabla \sigma(\X,t)\big], \\
    \label{eq:TDconseq}
    & \int_{\partial V_+(t)}dS_{\Y} \, \n(\Y)\cdot\vi(\Y,t) \Dbp(\Y,t) + \sum_{k=\pm}\int_{V_k(t)}d^3y \, \partial_t \bphi_k(\y,t) = 0~,
    \end{align}
\end{subequations}
where $\zeta(\X,\Y,t) = G_D(\X,\Y,t)\Dbp(\Y,t)$.
Equation~\eqref{eq:TDintegralmotion} exhibits a structure analogous to that of Eq.~\eqref{eq:integralmotion}, except that it includes the additional force $\Xi$.
Equation~\eqref{eq:TDconseq} results from the constraint of mass conservation, 
and also features an additional term with respect to Eq.~\eqref{eq:conseq}, that accounts for possible time modulations of the binodal concentrations. 
Importantly, like Eqs.~\eqref{eq:walleqs}, Eqs.~\eqref{eq:TDwalleqs} describe the interface dynamics only in terms of directly measurable quantities, such as the binodal concentrations, the chemical potential, the mobilities, and the surface tension.

We now illustrate a practical application of Eqs.~\eqref{eq:TDwalleqs} for a minimal model of spatio-temporally modulated mixture in one dimension, 
which we describe with the Ginzburg-Landau free energy
\begin{equation}
    \label{eq:TDGL}
    \F_{\rm GL}[\phi,t] = 
    \int dx \left[ 
    \frac{K}{2}|\partial_x\phi|^2+\frac{\tau(x,t)}{2}(\phi-\phi_c)^2+\frac{c}{4}(\phi-\phi_c)^4\right],
\end{equation}
where $\phi_c$ denotes the density at the critical point, 
while we keep the mobility constant: $D(\phi,\x,t) = D$.
The free energy~\eqref{eq:TDGL} is similar to that derived in Eq.~\eqref{eq:EffectiveGL} for a mixture subject to an inhomogeneous temperature profile near its critical point, except that we have neglected the linear contribution $\propto \phi - \phi_c$ for simplicity.
Equation~\eqref{eq:TDGL} is therefore relevant to mixtures presenting a weak Soret effect,
or for which the modulation $\tau(x,t)$ originates from other causes, 
like interactions with another chemical species \cite{jambon2023phase} or with light in the case of photosensitive biopolymers~\cite{Bracha_2018}. 

We focus on the case where the sample is subject to a moving source of chemical or light, such that we write $\tau(x,t)=\tau[x-x_s(t)]$, where $x_s(t)$ parametrizes the trajectory of the source.
As in Sec.~\ref{sec:1d_simple}, a droplet of concentration $\bphi_+$ is defined by the locations of its left and right interfaces, which we denote as $x_r$ and $x_l$, respectively. 
In the case described by Eq.~\eqref{eq:TDGL} where the binodal concentrations are dynamical quantities, however, the size of the droplet can change over time,
such that its dynamics cannot be expressed solely in terms of its centre of mass velocity.

Considering a periodic system of size $L$, Eqs.~\eqref{eq:TDwalleqs} then lead to:
\begin{equation} \label{eq:TDtermophoresis}
    \dot{x}_I = - \sum_{J=l,r} M_{{\rm eff},IJ}\partial_{x_{J}} U_{\rm eff} + \left(1 - \frac{\Omega}{\Dbp(x_I,t)}\right) \dot{x}_s~,
\end{equation}
where $I=l$ or $r$, $S(t)=|x_r(t)-x_l(t)|$ is the size of the droplet,
$\bphi_\pm(x,t) = \phi_c \pm \sqrt{-\tau(x,t)/c}$, while
\begin{align*}
    {M}_{{\rm eff},IJ}(x_r,x_l,t) & = \frac{D L}{S(t)[L-S(t)]}[\Dbp(x_I,t)\Dbp(x_J,t)]^{-1},\\
    U_{\rm eff}(x_r,x_l,t) & = \sigma(x_r,t) + \sigma(x_l,t),\\
    \Omega(x_r,x_l,t) & = \int_{V_+(t)}dx\, \frac{\bphi_+(x,t)}{S(t)} - \int_{V_-(t)}dx\, \frac{\bphi_-(x,t)}{L-S(t)}.
\end{align*}
The first term on the r.h.s.\ of Eq.~\eqref{eq:TDtermophoresis} exhibits a similar form as the droplet velocity appearing in Eq.~\eqref{eq:simpleEoM}.
Yet, we note that the surface tension ---and therefore the effective potential--- are now time-dependent, such that the resulting force is not conservative.
The second term on the r.h.s.\ of Eq.~\eqref{eq:TDtermophoresis} indicates that the droplet dynamics does not only depend on the position of the source, but also on its velocity.
Physically, this contribution accounts for mass conservation, which imposes that a droplet may shrink or expand as it experiences a dynamic landscape of binodal densities.
Indeed, this term cancels when the moving source does not affect the values of the binodals, as can be seen from the fact that $\Omega = \Delta\bphi$ when $\bphi_+$ and $\bphi_-$ are uniform in space. 

 \begin{figure}[!t]
    \centering
    \includegraphics[width=1.0\linewidth]{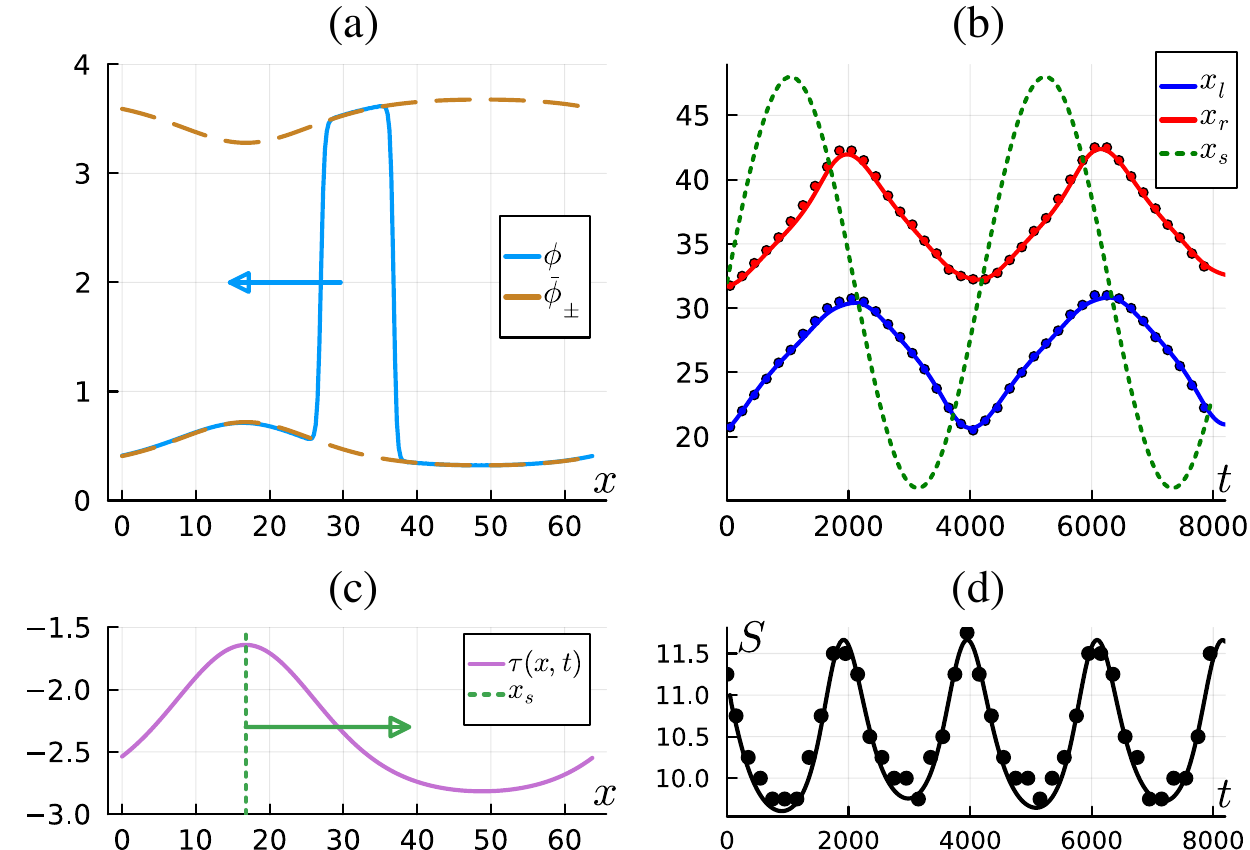}
    \caption{Numerical simulations of a conserved dynamics with free energy \eqref{eq:TDGL}.
    Panel (a): Shape of the droplet $\phi(x,t)$.  The yellow dashed line denotes the binodals, and the green dotted line the position of the source. Panel (b): temperature modulation $\tau(x,t)$. Blue and green arrows indicate the velocity of the droplet centre and source respectively. Panel (b):  position of the two interfaces. Numerical (points) against theoretical (lines) for the left and right interfaces. The green dotted line denotes the position of the source. Panel (d): size $S$ of the droplet as a function of time. Points are from simulations, lines from the interface dynamics.}
    \label{fig:temp_dep}
\end{figure}

We verify the validity of Eq.~\eqref{eq:TDtermophoresis} by performing numerical simulations of the full field equation~\eqref{eq:Current} with the free energy~\eqref{eq:TDGL}, 
and where $\tau < 0$ in the whole sample (Details on numerical simulations can be found in Sec. D of ESI$\dag$).
We consider the case where the moving source locally raises the value of $\tau$ (Fig.~\ref{fig:temp_dep}(c)), 
while its motion follows a sinusoidal trajectory.
Comparing the red and blue symbols with the  corresponding lines in Fig.~\ref{fig:temp_dep}(b) reveals excellent agreement between the simulated droplet trajectory and the predictions from Eq.~\eqref{eq:TDtermophoresis}.
Namely, we find that the droplet is attracted by the moving source, while the concentrations of both phases are well-approximated by the corresponding binodals (Fig.~\ref{fig:temp_dep}(a)).
As a result of the spatio-temporal modulation of the binodal concentrations, the motion of the droplet is also associated with changes of its shape $S(t)$, as shown in Fig.~\ref{fig:temp_dep}(d).
Namely, $S(t)$ is maximal when the source is located at the centre of the droplet,
which corresponds to the configuration for which the droplet concentration $\bphi_+$ is lowest.
The swelling of the droplet is then a simple consequence of the constraint of mass conservation. 
 
\section{Discussion}
\label{Discussion}

In this work, we have developed a general framework for describing the dynamics of interfaces in phase-separating systems that are subject to space- and time-dependent modulations of their material properties. 
The equations of motion, Eqs.~\eqref{eq:walleqs} and~\eqref{eq:TDwalleqs}, derived in the sharp interface approximation, allow us to quantify the dynamics of interfaces in terms of directly measurable quantities such as surface tension, chemical potential, binodal concentrations, and mixture mobility. 
We have shown that, even in the absence of hydrodynamics, the motion of phase-separated droplets is driven by two contributions of distinct physical origins. 
The first one is a capillary term, which causes droplets to minimize the overall surface tension at their interface, while the second results from bulk currents induced by a spatially-dependent chemical potential, which drive dense droplets to regions of minimum free energy. Notably, the relative magnitude of these terms is controlled by the droplet volume, with the capillary contribution dominating the dynamics of small droplets.

While we have illustrated the consequences of these results through minimal examples mostly involving a single droplet, Eqs.~\eqref{eq:walleqs} and~\eqref{eq:TDwalleqs} can potentially describe a wide range of physical phenomena. Considering multiple droplets will, for example, enable the exploration of how an inhomogeneous environment affects the properties of nucleation~\cite{debenedetti2020,Oxtoby_1992} and ripening~\cite{Bray_2002,rosowski2020elastic,cates_tjhung_2018}. Beyond inhomogeneities induced by temperature modulations, our approach is suited to investigate how the dynamics of droplets is influenced by other factors such as gravity~\cite{Shiwa1988}, electric fields~\cite{HasePRE2006}, and diffusiophoresis~\cite{golestanian2022a,Weber2017,AgudoPRL2019,Demarchi_2023}. 
In particular, a number of works have postulated that the positioning of biomolecular condensates in cells can be driven by diffusiophoresis in response to the presence of biomolecule concentration gradients~\cite{SearPRL2019,Hafner2024ACSNano}. Importantly, current studies have overlooked the potential impact of effective capillary forces induced by gradients of surface tension on the dynamics of these small droplets, whose diameter often ranges below the micrometer. 

The framework we have introduced here neglects hydrodynamics, such that Eqs.~\eqref{eq:walleqs} and~\eqref{eq:TDwalleqs} typically describe the motion of slow interfaces in a highly viscous environment. Extensions that take into account the coupling of phase separation with hydrodynamic flows, however, should be straightforward, provided that the typical flow length scale remains much larger than the interface width. 
Our approach further offers promising avenues for the study of active phase separation~\cite{AgudoPRL2019,golestanian2019,Saha2020,Cates_2025} in inhomogeneous media by considering, for example, active agents such as enzymes or bacteria interacting via self-produced chemical fields~\cite{AgudoPRL2019,Demarchi_2023,ZwickerNP2017,golestanian2017,DuanPRL2023}.
\section*{Conflicts of interest}
There are no conflicts to declare.
\bibliography{RR_bib,Golestanian} 
\bibliographystyle{rsc}


\newpage
\onecolumn

\begin{center}
    \LARGE{Dynamics of Phase-Separated Interfaces in Inhomogenous and Driven Mixtures \\
    ---Supplementary Information---}\\
    \vspace{2mm}
    \large{Jacopo Romano\textit{$^{a}$}, 
           Ramin Golestanian\textit{$^{a,b}$} and
           Beno\^it Mahault\textit{$^{a}$}}\\
    \large{\textit{$^{a}$~Max Planck Institute for Dynamics and Self-Organization (MPI-DS), 37077 G\"ottingen, Germany.}}\\
    \large{\textit{$^{b}$~Rudolf Peierls Centre for Theoretical Physics, University of Oxford, Oxford OX1 3PU, United Kingdom.}}\\
\end{center}

\renewcommand{\thesection}{\Alph{section}}
\setcounter{section}{0}
\renewcommand{\theequation}{S\arabic{equation}}
\setcounter{equation}{0}
\renewcommand{\thefigure}{S\arabic{figure}}
\setcounter{figure}{0}

This document provides calculation details supporting the results presented in the main text. It also includes information on the numerical simulations of the Cahn-Hilliard model, which we used to generate the results shown in Figs. 1 and 3 of the main text.

\section{The surface tension of an inhomogeneous mixture}

In this section, we outline the calculation steps leading to Eq.~(4) of the main text, which gives the expression of the surface tension for an inhomogeneous mixture in the sharp interface limit.
Here, we work in the regime where the interface is sufficiently thin such that we neglect its curvature and locally approximate it as flat.

First, let us consider a homogeneous mixture, i.e. for which the coefficients in the free energy do not explicitly depend on space.
Supposing that the interface is orthogonal to the $x$ axis, the corresponding concentration profile $\bphi(x)$ minimizes the free energy density 
\begin{equation}
\label{eqapp:energydens}
    \tilFdens(\phi)
    =\frac{K(\phi)}{2}\left(\frac{d\phi}{dx}\right)^2 + U(\phi) - \bar{\mu}\phi
    = \frac{K(\phi)}{2}\left(\frac{d\phi}{dx}\right)^2+\tilU(\phi),
\end{equation}
where we have kept the notations used in Sec. 2.1 of the main text. 
In particular, the chemical potential $\bar{\mu}$ enters as a Lagrange multiplier ensuring total mass conservation, while the profile $\bphi(x)$ satisfies the boundary conditions $\lim_{x\to \pm \infty} \bphi(x) = \bphi_{\pm}$ with $\bphi_\pm$ the binodal concentrations of the two phases. 
To determine $\bphi$, we thus apply the Euler-Lagrange equations to $F'$, leading to
\begin{equation}
    \label{eqapp:EoM1D}
    K(\bphi)\frac{d^2\bphi}{dx^2}+\dfrac{1}{2} K'(\bphi)\left(\frac{d\bphi}{dx}\right)^2-\tilU'(\bphi) = 0.
\end{equation}
We then multiply both sides of Eq.~\eqref{eqapp:EoM1D} by $d\bphi/dx$, and integrate from $-\infty$ to $x$, to obtain
\begin{equation}
    \label{eqapp:eomIdentity}
    \frac{1}{2} K(\bphi)\left(\frac{d\bphi}{dx}\right)^2 = \tilU(\bphi)-\tilU(\bphi_-),
\end{equation}
where we have used that $\bphi \to \bphi_-$ and $d\bphi/dx \to 0$ when $x \to -\infty$. 
The surface tension of the interface corresponds to the free energy of the system per unit area relative to an equilibrium homogeneous configuration. 
Using, moreover, the phase ``$-$'' as a reference, we obtain~\cite{Cahn_1958}
\begin{equation}
    \label{eqapp:intfreenexpr}
    \sigma = \int_{-\infty}^{+\infty} dx \; \left[ \frac{K(\bphi)}{2}\left(\frac{d\bphi}{dx}\right)^2+ \tilU(\bphi)-\tilU(\bphi_-) \right] 
    = \int_{-\infty}^{\infty} dx\; K(\bphi)\left(\frac{d\bphi}{dx}\right)^2
    = \int_{\bphi_-}^{\bphi_+} d\bphi \; K(\bphi)\frac{d\bphi}{dx},
\end{equation}
where the second equality results from Eq.~\eqref{eqapp:eomIdentity}, while the last one was obtained through the change of integration variable $x \to \bphi(x)$.
Solving Eq.~\eqref{eqapp:eomIdentity} for $d\bphi/dx$, and plugging the corresponding solution in Eq.~\eqref{eqapp:intfreenexpr}, we then get
\begin{equation}
    \label{eqapp:surftens}
    \sigma=\int_{\bphi_-}^{\bphi_+}d\bphi \; \sqrt{2K(\bphi)[\tilU(\bphi)-\tilU(\bphi_-)]}.
\end{equation} 

Assuming now that the coefficients in the free energy~\eqref{eqapp:energydens} vary in space,
the surface tension becomes a spatially-dependent quantity.
In the sharp interface limit, the coefficients $K(\phi,x)$, $\tilU(\phi,x)$ and $\bphi_\pm(x)$ entering the expression of $\sigma$ in Eq.~\eqref{eqapp:surftens} 
do not vary much over the interface thickness $l$. 
We can then simply approximate these coefficients with their value at the centre of the interface, which allows to straightforwardly recover Eq. (4) of the main text.

\section{Derivation of the interface equations of motion}
\label{app:Interface_Derivation}

In this section, we detail the calculation steps leading to the interface equations of motion for spatially modulated mixtures (Eq.~(9a) of the main text). 

\begin{figure}[t]
    \centering
    \includegraphics[width=.5\linewidth]{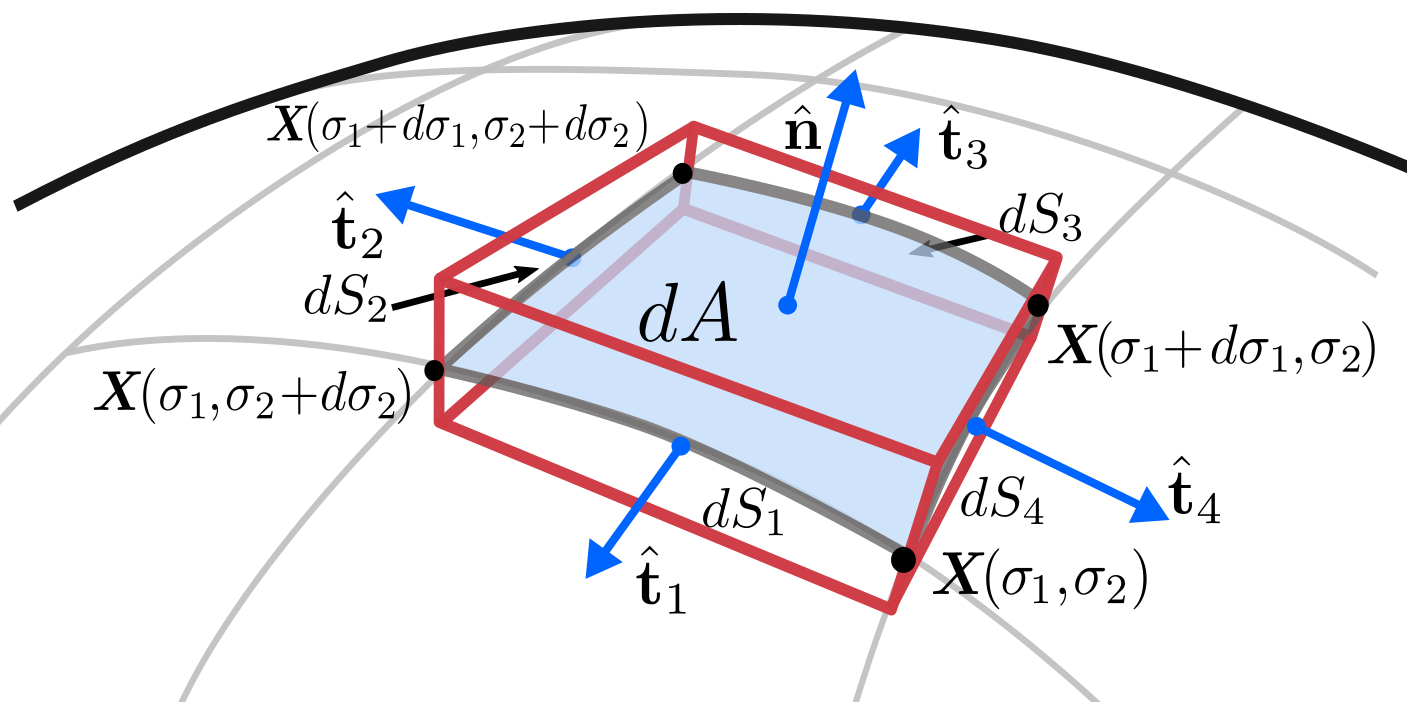}
    \caption{Schematic of the volume $dW$, marked by the red edges, which encloses an infinitesimal surface $dA$ of the interface.}
    \label{fig:variationdomainV}
\end{figure}

Hereafter, $\X$ denotes the coordinates of an arbitrary point of the interface, while $\n(\X)$ is the corresponding normal unit vector.
As explained in the main text, the interface equation of motion is obtained by multiplying the equation
\begin{equation}
    \label{eq:invcontinuity_SM}
    \int d^3y \; G_D(\X,\y)\,\partial_t\phi(\y,t)=\dfrac{\delta \tilF}{\delta\phi}(\phi,\X)+\comtg(\X)-\mu_{\infty},
\end{equation}
by $\n(\X) \cdot \nabla \phi$ and integrating over an infinitesimal volume $dW$ around $\X$.
As sketched in Fig.~\ref{fig:variationdomainV} for a two-dimensional interface, $dW$ encompasses an infinitesimal surface $dA$ of the interface, 
while its lateral faces orthogonal to $\n(\X)$ sit well into the bulk phases.
On the other hand, 
the dimensions of $dW$ in the remaining directions are assumed small compared to the radii of curvature of the interface, which we locally approximate as flat.

The integration of the second term on the r.h.s.\ of Eq.~\eqref{eq:invcontinuity_SM} is straightforward. 
Indeed, since $\comtg(\x)$ remains nearly constant across the interface,  we approximate $\comtg(\x) \simeq \comtg(\X)$ inside $dW$, and get
\begin{equation} \label{eq:int_mu_SM}
 \int_{dW} d^3x \; \n(\X)\cdot\nabla\phi(\x)[\comtg(\X)-\mu_{\infty}] = -dA \Dbp(\X) [\comtg(\X)-\mu_{\infty}],
\end{equation}
where we recall that $\Dbp(\X) = \bphi_+(\X) - \bphi_-(\X)$,
while the minus sign on the r.h.s.\ results from the fact that in the convention we use, $\n(\X)$ points from the phase ``$+$'' to the phase ``$-$''.
Similarly, and as detailed in the main text, integrating the l.h.s.\ of Eq.~\eqref{eq:invcontinuity_SM} yields
\begin{equation} \label{eq:int_lhs_SM}
    \int_{dW} d^3x \; \n(\X)\cdot\nabla\phi(\x) \int d^3y \; G_D(\x,\y)\,\partial_t\phi(\y,t) = -dA \Dbp(\X) \int_{\partial V_{+}(t)} d S_{\Y} \, \n(\Y)
    \cdot\vi(\Y,t) \zeta(\X,\Y)~,
\end{equation}
where $\zeta(\X,\Y) = \Dbp(\Y)G_D(\X,\Y)$.

The integration of the free energy functional derivative in Eq.~\eqref{eq:invcontinuity_SM} is, on the other hand, less straightforward.
A key identity for this purpose is (using implicit summation over repeated indices):
\begin{equation}
\label{eqapp:stress-consumption}
  \dfrac{\delta \tilF}{\delta \phi}(\phi,\x)\partial_j \phi(\x)=-\partial_i T_{ij}(\phi,\x)-\bar\partial_j \tilFdens(\phi,\x),
\end{equation}
where $T_{ij}=(\partial_{j}\phi) \frac{\partial \tilFdens}{\partial (\partial_{i} \phi)} - \delta_{ij} \tilFdens$ is the stress tensor associated to the free energy $\tilFdens$, 
while the notation $\bar\partial_i \tilFdens(\phi,\x)$ indicates that the derivative is taken with respect to the explicit spatial dependencies of $\tilFdens$ in the coordinates, 
and not on the ones originating through $\phi(\x)$.
Furthermore, below it will be convenient to separate explicit spatial dependencies from the ones induced by $\phi$ by defining
\begin{equation}
    \tilFdens_{\y}(\phi) = \frac{K(\phi,\y)}{2}|\nabla\phi(\x)|^2 + \tilU(\phi(\x),\y),\qquad {\rm and} \qquad
    \tilF_{\y}[\phi] = \int d^3x\,\tilFdens_{\y}[\phi(\x)],
\end{equation}
such that, for every field $\psi(\x)$,
\begin{equation}
    \bar\partial_j \tilFdens[\psi(\x)] = \frac{\partial}{\partial y_j}\tilFdens_{\y}[\psi(\x)]\bigg|_{\x=\y}.
\end{equation}

We then define $\bphi_{\y}(\x)$ as the minimizer of $\tilF_{\y}$, i.e. the field solution of:
\begin{equation}
\label{eqapp:ELtilde}
    \left[\partial_i \left( \frac{\partial \tilFdens_{\y} }{\partial[\partial_i\phi(\x)]}\right)
    - \frac{\partial \tilFdens_{\y} }{\partial\phi(\x)}\right]_{\phi(\x) = \bphi_{\y}(\x)}=0.
\end{equation}
This way, $\tilde\phi_{\y}(\x)$ is nothing else but the interface profile, computed at some position $\x$ as if the spatially-dependent parameters of the theory were equal to their value at some other position $\y$. 

As we argue in the main text, in the sharp interface limit the interface profile relaxes to its equilibrium shape $\bphi$ over a vanishing time scale. 
In addition, since the interface is much thinner than any length scale associated with the variation of the model parameters, we have 
\begin{equation} \label{eqapp:rigid_core_approx}
    \phi(\x,t) = \bphi_{\x}[\x - \X(t)]+O(l),
\end{equation}
 which states that at leading order in the interface thickness $l$,
 the profile of the interface at some position $\X$ is given by the stationary profile $\bphi(\x)$ where the parameters of the theory are evaluated at $\X$.
Then, the surface tension is given by:
\begin{equation} \label{eqapp:sigma_def}
    \sigma(\x)=\int_{-\infty}^{\infty} dx_\perp \left[\dfrac{K(\phi,\x)}{2}(\partial_{x_\perp}\phi)^2+\tilU(\phi,\x) \right]
    =\int_{-\infty}^{\infty} dx_\perp \left[
    \frac{K(\bphi_{\y},\y)}{2}(\partial_{x_\perp}\bphi_{\y})^2+\tilU(\bphi_{\y},\y)dn\right]\at{\y=\x}+O(l)
    =\int_{-\infty}^{\infty}  dx_\perp \; F'_{\y}[\bphi_{\y}(\x)]\at{\y=\x}+O(l),
\end{equation}
where the $x_\perp$ variable indicates that the integration is performed over a line orthogonal to the interface.

We now proceed with the integration of Eq.~\eqref{eqapp:stress-consumption} in the volume $dW$.
The integration of the term on the r.h.s.\ deriving from the stress tensor has been presented in Ref.~\cite{Romano_2024, Romano2023}.
Using the divergence theorem, this contribution leads to a surface term, while the integrals over the faces of $dW$ normal to ${\n}$, which sit exclusively in the bulk phases, vanish, as $T_{ij}$ is identically zero in these phases.
Therefore, we consider the four remaining lateral sides, which we denote $dS_I$ with $I=1,2,3,4$ (see Fig.~\ref{fig:variationdomainV}), together with $\hat{\bm t}_I$ the corresponding normal unit vector and $\x_I$ their geometric centre. 
For convenience, we also locally parametrize the surface defining the interface by the coordinates $(\sigma_1,\sigma_2)$, such that: $\X=\X(\sigma_1,\sigma_2)$
and we work in an orthogonal parametrization such that $\partial_{\sigma_1}\X\cdot\partial_{\sigma_2}\X=0$.
Then, $dA = [\X(\sigma_1 + d\sigma_1,\sigma_2) - \X(\sigma_1,\sigma_2)]\cdot[\X(\sigma_1,\sigma_2 + d\sigma_2) - \X(\sigma_1,\sigma_2)]$, and
\begin{equation}
    \hat {\bm t}_1=-|\partial_{\sigma_1}\X|^{-1}\partial_{\sigma_1}\X|_{\x_1}, \qquad
    \hat{\bm t}_2=|\partial_{\sigma_2}\X|^{-1}\partial_{\sigma_2}\X|_{\x_2}, \qquad
    \hat{\bm t}_3=|\partial_{\sigma_1}\X|^{-1}\partial_{\sigma_1}\X|_{\x_3}, \qquad
    \hat{\bm t}_4=-|\partial_{\sigma_2}\X|^{-1}\partial_{\sigma_2}\X|_{\x_4}.
\end{equation}

Since the sides $dS_I$ are orthogonal to the interface and $|\hat{\bm t}_I\cdot\nabla\phi| \ll |\n\cdot\nabla\phi|$, we obtain for all $I$
\begin{equation}
    \label{eqapp:siteintegral}
    -\int_{dS_I} dS_i \; T_{ij}(\phi,\x) 
    = \hat{t}_{I,i}(\x_I) dX_I \int dx_\perp \; \tilFdens(\phi,\x_I)
    = \hat{t}_{I,i}(\x_I)\, dX_I\,\sigma(\x_I),
\end{equation}
where $dX_I$ is the lateral length of $dS_I$ and we have used Eqs.~\eqref{eqapp:rigid_core_approx} and~\eqref{eqapp:sigma_def} to get the last equality. 
Noting that, at leading order in $d\sigma_{1,2}$, $\hat{\bm t}_{1} + \hat{\bm t}_{3} = -dX_2 \kappa_1 \n$ and $\hat{\bm t}_{2} + \hat{\bm t}_{4} =  -dX_1 \kappa_2 \n$ with $\kappa_1$ and $\kappa_2$ the surface principal curvatures, we then obtain after summing over the four contributions 
\begin{equation}
    \label{eqapp:stresspart}
    -\hat n_j\int_{d W}dS_i T^{ij}(\phi,\x)=-2 \sigma(\X) H(\X)dA,
\end{equation}
where $H(\X)$ is the mean curvature of the interface evaluated at $\X$.
To integrate the second contribution on the r.h.s. of Eq.~\eqref{eqapp:stress-consumption}, we use the identity:
\begin{align}
\label{eqapp:identity}
    \bar\partial_{y_j}\tilFdens_{\y}[\bphi_{\y}(\x)] = \partial_{y_j} \tilFdens_{\y}[\bphi_{\y}(\x)]
    - \frac{\partial\tilFdens_{\y}}{\partial \phi}\partial_{y_j}\bphi_{\y}(\x)
    - \frac{\partial\tilFdens_{\y}}{\partial \partial_{x_i}\phi}\partial^2_{x_i y_j}\bphi_{\y}(\x)
    =\partial_{y_j} \tilFdens_{\y}[\bphi_{\y}(\x)]
    -\partial_{x_i}\left(\partial_{y_j}\bphi_{\y}(\x)\frac{\partial\tilFdens_{\y}}{\partial \partial_{x_i}\phi}[\bphi_{\y}(\x))]\right),
\end{align}
where we have used Eq.~\eqref{eqapp:ELtilde} to obtain the second equality, and as stated before $\bar\partial_y$ only acts on explicit $\y$ dependencies of $F'_{\y}$.
We then have, at the dominant order in $l$,
\begin{equation}
    \label{eqapp:stresspart2}
    -\hat n_j\int_{dW}d^3x \; \bar\partial_j \tilFdens(\phi,\x)
    = -\hat n_j\int_{dW}d^3x \; \partial_{y_j} \tilFdens_{\y}[\tilde \phi_{\y}(\x)]\at{\y=\x}
    = -\hat n_j\partial_{y_j}\left(\int_{dW}d^3x \; \tilFdens_{\y}[\tilde \phi_{\y}(\x)]\right)\at{\y=\X}=-\hat n_j\partial_{X_j}\sigma(\X)dA
\end{equation}
where we have used Eq.~\eqref{eqapp:identity} to obtain the second equality, together with the fact that integral of the divergence term vanishes.
Combining Eqs.~(\ref{eq:int_mu_SM}, \ref{eq:int_lhs_SM}, \ref{eqapp:siteintegral}, \ref{eqapp:stresspart2}), we then obtain Eq.~(9a) given in the main text, while the derivation of Eq.~(9b) follows a similar approach.

\subsection*{The time-dependent case}
We now briefly comment on the derivation of Eqs.~(38) for mixtures undergoing spatio-temporal modulations.
The above results can be straightforwardly generalized to the time-dependent coefficients by including an explicit time dependency in the binodal concentrations, common tangent, and surface tension in Eq. (9).
Having time-dependent binodal concentrations, however, leads to an additional term when calculating time derivative of $\phi$. 
Namely, replacing $\phi(\x,t)=\bphi_-(\x,t)+\Dbp(\x,t) \,\Theta[\x,V_+(t)]$ in the l.h.s.\ of Eq.~\eqref{eq:invcontinuity_SM}, we obtain
\begin{align}
    \label{eqapp:timechange}
    &\int d^3 y \; G_D(\X,\y,t) \partial_t\phi(\y,t) =
    \int_{\partial V_+(t)} dS_{\Y} \, \n(\Y)\cdot\vi(\Y,t)\, \zeta(\X,\Y,t)
    +\int_{V_+(t)} d^3y \; G_D(\X,\y,t) \, \partial_t\bphi_{+}(\y,t) + \int_{V_-(t)} d^3y \; G_D(\X,\y,t) \, \partial_t\bphi_{-}(\y,t),
\end{align}
where $\zeta(\X,\Y,t) = \Dbp(\Y,t)G_D(\X,\Y,t)$, while $V_+(t)$ and $V_-(t)$ denote the volumes of the phases with concentration $\bphi_+(t)$ and $\bphi_-(t)$, respectively, which are separated by the interface $\partial V_+(t)$.
As discussed above, since the parameters of the model do not vary much across the interface, we have
\begin{equation}
    \label{eqapp:timechange2}
  \int_{dW} d^3x \; \n(\x)\cdot\nabla\phi(\x,t) \int d^3 y \; G_D(\x,\y,t)\, \partial_t\phi(\y,t)
    = - dA \Dbp(\X,t) \left[ \int_{\partial V_+(t)} dS_{\Y} \n(\Y)\cdot\vi(\Y,t) \zeta(\X,\Y,t) - \Xi(\X,t) \right],
\end{equation}
where $\Xi(\X,t) = -\sum_{k=\pm}\int_{V_k(t)} d^3 y\; G_D(\X,\y,t) \, \partial_t\bphi_k(\y,t)$ as defined in the main text.
In addition, Eq.~(38b) of the main text is obtained from a similar calculation after replacing the ansatz for $\phi(\x,t)$ in the mass conservation constraint $\int d^3x \; \partial_t \phi(\x,t)$.

\section{Green's functions in periodic domains}
\label{Green_SM}

To confront the predictions of the interface equations of motion ---Eqs.~(9) and~(38) in the main text--- to numerical simulations of the full phase field model, 
we considered one-dimensional systems with periodic boundary conditions.
Here, we detail how we calculate the corresponding Green's function, which differs from that associated with infinite domains.

Namely, we consider a periodic domain of size $L$, and look for a function $G_D(x,y)$ satisfying the boundary conditions $G(0,y)=G(L,y)$ and $\partial_xG(0,y)=\partial_xG(L,y)$.
First, we note that the equation
\begin{equation}
    \label{eqapp:originalgreeneq}
    \partial_x \,[D(\phi,x)\,\partial_x G_D(x,y)]=\delta(x-y),
\end{equation}
does not admit such a periodic solution for $G_D(x,y)$, as can be checked by integrating it over $x$ in the whole domain.
To find an admissible Green's function, we instead look for solutions of 
\begin{equation}
    \label{eqapp:periodicgreeneq}
    \partial_x \,[D(\phi,x)\,\partial_x G_D(x,y)]=\delta(x-y)-\dfrac{1}{L}.
\end{equation}
The idea behind this approach is that the convolution
\begin{equation} \label{eqapp:psi}
    \psi(x) = \int_0^L dy \; G_D(x,y)s(y),
\end{equation}
with $G_D(x,y)$ a periodic solution of Eq.~\eqref{eqapp:periodicgreeneq}
is still a solution of the equation $\partial_x \,[D(\phi,x)\,\partial_x \psi(x)]=s(x)$ whenever the source satisfies $\int_0^L dx \; s(x) = 0$, 
as can be checked by replacing $\psi(x)$ with Eq.~\eqref{eqapp:psi}.
The phase field model with conserved mass precisely falls into the same class of problems, where $\psi$ is the mixture chemical potential, while $s = \partial_t \phi$ whose integral over the whole domain vanishes due to mass conservation.
Crucially, contrary to Eq.\eqref{eqapp:originalgreeneq}, Eq.~\eqref{eqapp:periodicgreeneq} admits a periodic solution, which for constant $D$ is given by
\begin{equation}
   \label{eqapp:periodicgreen}
   G_D(x-y)=\dfrac{|x-y|(L-|x-y|)}{2 D L}.
\end{equation}

\section{Details on numerical simulations}
\label{Simulations}

The data shown in Fig.~1 of the main text was obtained by simulating the phase field model, Eq. (1), with the free energy given in Eq. (10).
We have used as parameters $D=K=1$, $L=64$, 
$A(x)=10-2.5 \exp[-\cos(2\pi x/L)]$, $\bphi_+=2$, and $\bphi_-=0$.
The simulation was initialized with a rectangular droplet with interfaces sitting at $x_l=7.$ and $x_r=20.75$.

The data shown in Fig. 3 of the main text was obtained by simulating the phase field model with the free energy of Eq.~(39), featuring space and time modulations.
Here, we used $D=K=1$, $L=64$, $c=4$, and $\phi_c=2$, while the effect of the source was modeled by the quadratic coefficient
\begin{equation}
    \tau[x-x_s(t)]=-3+\frac{1}{2}
    \exp\left[-\cos\left(\frac{2\pi[x-x_s(t)]}{L}\right)\right],
\end{equation}
where $x_s(t)=\tfrac{1}{4}L \sin(v t)+\tfrac{1}{2}L$ and $v=1.5\times 10^{-3}$.

Simulations of the phase field model were performed by means of a pseudo-spectral method with an explicit Euler integrator for time update, using space and time resolutions $dx = \tfrac{1}{4}$ and $dt=5\times10^{-5}$, respectively. 
Simulations of the interface equations are performed by explicit Euler method with time resolution $dt=20$ and $dt=8$ for Fig.~1 and Fig.~3, respectively.

\section{Dynamics of spherical droplets subjected to surface tension gradients}
\label{app:dynamicsCoM}

Here, we provide details on the derivation of Eq.~(13) of the main text, which describes the dynamics of spherical droplets under the influence of surface tension gradients.
For simplicity, we assume that the droplet shape is not affected by the surface tension landscape, such that it has a constant density and retains its spherical shape at all times.
The dynamics of the droplet is therefore characterized by the motion of its centre of mass, $\x_c = R^{-d}V_d^{-1}\int_{V_R(\x_c)} d^dx \; \x$,
where the integration domain $V_R(\x_c)$ denotes the ball of radius $R$ centred in $\x_c$, and $V_d$ is the volume of the unit $d$-sphere.
Using Reynolds transport theorem, the centre of mass velocity follows
\begin{equation}
    \label{eqapp:vcom}
    \dot{\x}_c=\frac{1}{R^{d} V_d}\int_{S_R(\x_c)} dS_\X \; \n(\X)\cdot \vi(\X,t) \; R\n(\X) = \frac{1}{R^{d-1} V_d}\int_{S_R(\x_c)} dS_\X \; \vi(\X,t),
\end{equation}
where the integration is now performed on the sphere of radius $R$ centred in $\x_c$, $S_R(\x_c)$. 

We consider the simple case of an infinite domain, for which the Green's functions in two and three dimensions are given by
\begin{equation}
    G^{(2)}_D(\x-\y)=\frac{1}{2\pi D}\log(|\x-\y|), \qquad
    G^{(3)}_D(\x-\y)=-\frac{1}{4\pi D|\x-\y|},
\end{equation}
respectively.
In particular, these functions satisfy the identity
\begin{equation}
    \label{eqapp:greensid}
    \int_{S_R(\x_c)} dS_\X \; G^{(d)}(\X-\Y)\n(\X) = -\frac{R}{d D}\n(\Y),
\end{equation}
where $\X$ and $\Y$ are two points on the circle ($d=2$) or on the sphere ($d=3$).

Multiplying Eq.~(9a) of the main text by $\n(\X)$, and integrating over $S_R(\x_c)$, we then get for the l.h.s.
\begin{equation}
    \int_{S_R(\x_c)} dS_\X \, \n(\X) 
    \int_{S_R(\x_c)}dS_{\Y}\, \n(\Y)\cdot\vi(\Y,t) \Dbp G^{(d)}(\X-\Y) = -\frac{\Dbp R^d V_d}{d D}\dot{\x}_c ,
\end{equation}
where we have used that the binodal concentrations are constant in the case we are interested in.
For the r.h.s.\ of Eq.~(9a), we note that the chemical potential is constant and can thus be discarded, such that we are left with evaluating the r.h.s.\ of
\begin{equation}
\label{eqapp:prelCoM}
    \frac{\Dbp^2 R^d V_d}{d D}\dot{\x}_c = -\int_{S_R(\x_c)}dS_\X \; \n(\X)\left[\frac{d-1}{R}\sigma(\X)+\hat{n}(\X)\cdot \nabla\sigma(\X)\right].
\end{equation}
Using the relation $\frac{(d-1)}{R}\sigma(\X)+\n(\X)\cdot \nabla\sigma(\X)= \nabla\cdot[\n(\X)\sigma(\X)]$, and applying the divergence theorem separately on each component of the r.h.s.\ of Eq.~\eqref{eqapp:prelCoM}, we obtain
\begin{align} \label{eqapp:spherial_drop_rhs}
\int_{S_R(\x_c)}dS_\X \; \hat{n}_i \left(\frac{d-1}{R}\sigma(\X)+\n(\X)\cdot \nabla\sigma(\X)\right)
& = \int_{V_R(\x_c)} dV \; \partial^2_{\X_i\X_j}[\hat{n}_j(\X)\sigma(\X)] \nonumber \\
& = \int_{V_R(\x_c)} dV \; \partial_{\X_j}\left[\hat{n}_j(\X)\partial_{\X_i}\sigma(\X) + \frac{\delta_{ij} - \hat{n}_i(\X)\hat{n}_j(\X)}{R}\sigma(\X)\right] \nonumber \\
& = \int_{S_R(\x_c)}dS_\X \; \partial_{\X_i}\sigma(\X),
\end{align}
where only the gradient term remains in the last equality, since the other term is orthogonal to $\n$.
To obtain Eq.~(13) of the main text, we finally use the fact that the droplet has a fixed shape, such that applying the change of variables $\Z=\X+\x_c$ in the integral on the r.h.s.\ of Eq.~\eqref{eqapp:spherial_drop_rhs} yields
\begin{equation}
    \int_{S_R(\x_c)} dS_{\X} \; \nabla\sigma(\X)=\int_{S_R(0)}dS_{\Z}  \nabla_{\Z}\sigma(\Z+\x_c)= \nabla_{\x_c}\int_{S_R(0)} dS_{\Z} \sigma(\Z+\x_c)=\nabla_{\x_c}\int_{S_R(\x_c)} dS_{\X} \,\sigma(\X).
\end{equation}
\end{document}